\RequirePackage{ifpdf}
\input epsf.tex
\epsfclipon
\documentclass[11pt,a4paper]{article}
\usepackage{jheppub}
\usepackage[numbers,sort&compress]{natbib}
\usepackage{comment}
\usepackage{epsfig,multicol}
\usepackage{adjustbox}
\usepackage{longtable}
\usepackage{amsmath} 
\usepackage{float}
\usepackage{hyperref}
\usepackage{doi}
\usepackage{longtable} 
\usepackage{graphicx}
\graphicspath{/Figures}
\usepackage{multirow}
\usepackage{float}
\usepackage[T1]{fontenc} 
\usepackage{cancel}
\usepackage[framemethod=TikZ]{mdframed}
\usepackage{lipsum}
\allowdisplaybreaks
\usepackage{stackengine}
\usepackage{tikz}
\usepackage{soul}
\usepackage{longtable}
\usepackage{makecell}
\usepackage{booktabs}
\usepackage{array}
\renewcommand\arraystretch{1.1}

\usetikzlibrary{%
  arrows,
  knots,
  calc,
}

\usepackage[utf8]{inputenc}
 
\usepackage{ytableau}
 \ytableausetup{centertableaux,smalltableaux}
 
\usepackage[mathscr]{euscript}

\stackMath

\usepackage{hyperref}
\usepackage{multirow,tabularx}

\usepackage{cleveref}
\usepackage{mathtools}
\usepackage{longtable}
\usepackage{stackengine}
\usepackage{color}

\usepackage{epsfig}
\usepackage{epstopdf}
\usepackage{epsf}
\usepackage{ytableau}
 \ytableausetup{centertableaux}
 \usepackage{tikz}
\usetikzlibrary{
  knots,
  hobby,
  decorations.pathreplacing,
  shapes.geometric,
  calc
}

\tikzset{
  knot diagram/every strand/.append style={
    ultra thick,
    red
  },
  show curve controls/.style={
    postaction=decorate,
    decoration={show path construction,
      curveto code={
        \draw [blue, dashed]
        (\tikzinputsegmentfirst) -- (\tikzinputsegmentsupporta)
        node [at end, draw, solid, red, inner sep=2pt]{};
        \draw [blue, dashed]
        (\tikzinputsegmentsupportb) -- (\tikzinputsegmentlast)
        node [at start, draw, solid, red, inner sep=2pt]{}
        node [at end, fill, blue, ellipse, inner sep=2pt]{}
        ;
      }
    }
  },
  show curve endpoints/.style={
    postaction=decorate,
    decoration={show path construction,
      curveto code={
        \node [fill, blue, ellipse, inner sep=2pt] at (\tikzinputsegmentlast) {}
        ;
      }
    }
  }
}

\input{epsf.sty}
\usepackage{graphicx,amsmath,amssymb}
\usepackage{graphicx}
\usepackage{epsfig,multicol}
\usepackage{multirow}
\allowdisplaybreaks
\hypersetup{
  colorlinks=true,
  citecolor=magenta,
  linkcolor=blue,
  urlcolor=violet
 }
\usepackage{tikz} 

\definecolor{mycolor}{rgb}{0.122, 0.435, 0.698}

\newmdenv[innerlinewidth=0.5pt, roundcorner=4pt,linecolor=mycolor,innerleftmargin=6pt,
innerrightmargin=6pt,innertopmargin=6pt,innerbottommargin=6pt]{mybox}

\usepackage{bbm,bm,amsmath,amssymb}
 
\usepackage[normalem]{ulem}

\usepackage{comment}
\def\be{\begin{equation}}
\def\ee{\end{equation}}
\def\figs/B{B}
\def\bea{\begin{eqnarray}}
\def\eea{\end{eqnarray}}
\def\bg{\begin{eqnarray}}
\def\nd{\end{eqnarray}}
\def\sin{{\rm sin}}

\def\log{{\rm log}}

\renewcommand{\d}{{\mathrm{d}}}
\renewcommand{\[}{\left[}

\def\be{\begin{equation}}
\def\ee{\end{equation}}

\def\doi{http://doi.org}

\def\d{\mathrm{d}}

\def\XXint#1#2#3{{\setbox0=\hbox{$#1{#2#3}{\int}$ }
\vcenter{\hbox{$#2#3$ }}\kern-.6\wd0}}

\usepackage{physics}
\usepackage{tikz}
\usetikzlibrary{arrows.meta}
\usetikzlibrary{angles,quotes} 
\usetikzlibrary{decorations.markings,arrows.meta}
\tikzset{>=latex} 

\tikzset{
  midarr/.style={decoration={markings,mark=at position #1 with {\arrow{stealth}}},postaction={decorate}},
  midarr/.default=0.5
}
\colorlet{xcol}{blue!70!black}

\title{Flat connections at infinity on knot surgery manifolds}

\author[a]{Aditya Dwivedi,}
\author[a]{Archana Maji,}
\author[b]{Dmitry Noshchenko,}
\author[a]{and Ramadevi Pichai}

\affiliation[a]{Department of Physics, Indian Institute of Technology Bombay,\\ Powai, Mumbai 400076, India}
\affiliation[b]{School of Theoretical Physics, Dublin Institute for Advanced Studies,\\ 10 Burlington Road, Dublin, D04 C932, Ireland}

\emailAdd{aditya.dwivedi@iitb.ac.in}
\emailAdd{archana\_phy@iitb.ac.in}
\emailAdd{dsnoshchenko@stp.dias.ie}
\emailAdd{ramadevi@iitb.ac.in}

\abstract{$\rm SL(2,\mathbb{C})$ Chern-Simons theory on a closed 3-manifold is one of the most interesting, yet tractable examples of a QFT. On one hand, its non-perturbative aspects are not yet fully understood; on the other, the mathematical structure turns out to be very rich. In this work, we explore the new phenomenon of flat connections at infinity on various knot surgery manifolds, continuing the line of research of \cite{gukov2024categorification}. Such flat connections can be understood as asymptotic ends in the non-compact moduli space of all $\rm SL(2,\mathbb{C})$ connections, suggesting that they may contribute to the Borel resummation of the perturbative Chern-Simons partition function. We focus on the examples of $\pm 1/r$-surgeries on torus, twist and some double twist knot complements in $S^3$. Surprisingly, our findings confirm the existence of flat connections at infinity even for simple low-crossing knot surgeries. We therefore believe that their presence would shed light on the resurgent nature of the path integral.}

\begin{document} 
\maketitle
\flushbottom

\section{Introduction}

In quantum theory, perturbation series often give the only feasible way to probe the underlying physics when the exact methods are not readily available. Thanks to the beautiful theory of resurgence by J. Écalle \cite{Ec81}, the perturbative data can serve as a window to the non-perturbative world by means of the Borel resummation. The idea of resurgence maintains a lot of research momentum in recent years. Mathematicians and physicists apply it in various case studies, which yielded some impressive results \cite{Dunne:2016nmc,ABS19,AM22,Baj21,Marino:2023epd,Adams:2025qgj}. 
For us, the relevant case is the Chern-Simons theory – a kind of topological quantum field theory defined on a 3-manifold with gauge group $\rm G$. In our work we focus on the choice $\rm G = SL(2,\mathbb{C})$\footnote{We use the terms complex Chern-Simons theory and $\rm SL(2,\mathbb{C})$ Chern-Simons theory interchangeably.}.
With this choice of the gauge group, the Chern-Simons theory is notably known for its connections to M-theory/quantum gravity, as well as the low-dimensional topology via the 3d-3d correspondence \cite{Gukov:2003na,Dimofte:2011ju,Chung:2014qpa,Chun:2019mal,Cheng:2022rqr,Gukov:2023cog}. 
It is also an example
where the perturbative calculations are available to a relatively high order (in comparison, for example, with QCD), which allows for a successful application of resurgence \cite{Gukov:2016njj,Costin:2023kla,gukov2024categorification,Harichurn:2025suf,Adams:2025aad}.

This work focuses on one special feature of complex Chern-Simons theory, namely, the non-compactness of its gauge group. About a decade ago, resurgent techniques were applied to the perturbative expansion of the Chern-Simons partition function on a closed 3-manifold $M$. This resulted in a collection of new $q$-series invariants, denoted $\widehat{Z}_a$, labelled by abelian flat connections on $M$ \cite{Gukov:2016njj,Gukov:2017kmk}.

One important structural property expected from $\widehat{Z}_a$ on a general 3-manifold is the decomposition in terms of the saddles of the $\rm SL(2,\mathbb{C})$ Chern-Simons functional, which are complex flat connections.
However, as was first noted in \cite{gukov2024categorification}, since the gauge group is non-compact, so is the moduli space of flat connections on $M$. Consequently, it may have infinite asymptotic ends -- infinite sequences of non-flat connections having a singular limit with vanishing curvature.
We refer to them as \emph{flat connections at infinity}.\footnote{The authors of \cite{gukov2024categorification} also use the term ``black swan'' to signify their unexpected appearance.}
This may result in extra contributions to $\widehat{Z}_a$, which do not correspond to the usual saddles of the path integral. (Among physicists, closely related objects that appear in other instances of QFT, are known as renormalons \cite{Dunne:2012ae, Shifman:2014fra, Boito:2021ulm, Ma21, Reis:2022tni}.)
Note that our case study naturally extends the phenomena of critical points at infinity (\cite{bahri1995variational, durfee1998five, bahri2006critical}; see also \cite{behtash2018critical})
into the infinite-dimensional setting of path integrals.
All the above said serves as a broad motivation for our work. 

\paragraph{} The pioneering study of flat connections at infinity \cite{gukov2024categorification} focused on a very special class of 3-manifolds, namely, $\pm 1$-surgeries on the complement of high-crossing prime knots and composite knots. Amongst them, there is direct evidence that such flat connections contribute to the Borel resummation of the perturbation series, and therefore are an integral part of the non-perturbative picture.
The main goal of this paper is to extend this direction further to suggest that flat connections at infinity are abundant for 3-manifolds coming from rational surgeries even on simple low crossing knots, such as torus and twist knots. 

We approach this problem by combining the Wirtinger presentation of the fundamental group of a knot with a particular ansatz for a flat connection at infinity. On the other hand, by analysing the zero locus of the knot $A$-polynomial along with the rational surgery constraint, we get another, independent confirmation on the number of such flat connections.
Our case studies include $\pm 1/r$-surgeries on torus, twist and double-twist knot families. For example, the number of {flat connections at infinity} for $\pm 1/2r$-surgery on twist knot $K_n$ grows with the number of twists $n$, as shown in Table \ref{table 1}.

\begin{table}[t]
\centering
\begin{adjustbox}{valign=c} 
\begin{tabular}{||c||c|c|c|c|c|c|c|c||}
\hline
   {$n$} & $-4$ & $-3$ & $-2$ & $-1$ &  $1$ & $2$ & $3$ & $4$ \\ \hline
$K_n$ & $10_1$ & $8_1$ & $6_1$ & $4_1$  & $3_1$ & $5_2$ & $7_2$ & $9_2$ \\ \hline
\text{\# flat conn. at $\infty$}&8&6&4&2&1&3&5&7\\ \hline
\end{tabular}
\end{adjustbox}
\caption{Twist knots $K_n$ and the number of flat connections at infinity on $S^3_{\pm 1/2r}(K_n)$} 
\label{table 1}
\end{table}

For a class of Brieskorn spheres coming from rational surgeries on a torus knot, we find an interesting branching pattern between the flat connections on a knot complement and those on the corresponding surgery.
This suggests a refinement of the $\rm SL(2,\mathbb{C})$ Casson invariant for Brieskorn sphere $\Sigma(s,t,rst\pm 1)$ which takes into account flat connections at infinity (see Section \ref{sec: counting} for the details).
Brieskorn spheres are special manifolds for which there is an equivalent description involving plumbed graphs \cite{gukov2021two}.
Whenever applicable, we reconcile our results with the plumbed graph method based on the adjacency matrix associated with such graphs (details of this method can be found in Appendix \ref{plumbed methd}).
Our findings may serve as the initial step toward a systematic understanding of the non-perturbative effects in complex Chern-Simons theory on a general 3-manifold.

The paper is organized as follows:

\begin{itemize}
    \item Section \ref{sec:prereq} familiarises the reader with the necessary material -- including $\rm SL(2,\mathbb{C})$ Chern-Simons theory, resurgence, rational surgeries on knots, flat connections and their Chern-Simons invariants.
    \item Section \ref{sec: counting} details the two approaches to counting flat connections at infinity on a surgery manifold. The first count is derived from the analysis of the fundamental group presentation, and is fully general. The second is conjecturally related to the degree of the map from the moduli space of flat connections to complex numbers, and in the present form is applicable to all Brieskorn spheres.
    \item Section \ref{sec: case studies} contains our main case studies and examples.
    First, we provide a detailed analysis of a flat connection at infinity for $-1/2$-surgery on the trefoil knot.
    Second, we move on to broader studies of classes of torus and twist knots, as well as some double-twist knots. We tabulate all flat connections at infinity and also find their Chern-Simons invariants by using both approaches from Section \ref{sec: counting}.
    \item Section \ref{sec:conclusions} is the summary of our results, as well as suggestions for future research directions.
\end{itemize}

\section*{Acknowledgements}

The authors~would like to thank Pavel Putrov for useful comments and advice.
The authors~would also like to acknowledge the SPARC/2019-2020/ P2116/ project funding for the 4 day workshop ``Knots, Quivers and Beyond'' in February 2025 where all of them could meet for discussion on this work. The work of AM is supported in part by the Prime Minister’s Research Fellowship provided by the Ministry of Education, Government of India.

\section{Prerequisites}
\label{sec:prereq}

We begin with recollecting some of the material required for our main studies. This includes complex Chern-Simons theory, the fundamentals of resurgence and moduli spaces of flat connections on a 3-manifold. We stress that our presentation here is very expository, and refer the reader to the original sources \cite{Gukov:2003na,Gukov:2016njj,Costin:2023kla,gukov2024categorification} for more details.

\paragraph{$\rm SL(2,\mathbb{C})$ Chern-Simons theory.}
\begin{enumerate}
    \item Let $M$ be an oriented three-dimensional manifold\footnote{In our work we primarily consider the two cases:
    (i)
    $M=S^3\setminus K$, complement of knot $K$ in three-sphere $S^3$  and (ii) $M=S^3_{\pm1/r}(K)$, a $\pm1/r$-surgery on a knot $K$ in $S^3$.} and $\rm G=\rm SL(2,\mathbb{C})$ a gauge group.
\end{enumerate}
Denoting $\mathcal{M}$ the space of connection 1-forms $A\in \Omega^1(M)\otimes\mathfrak{sl}(2,\mathbb{C})$ on $M$ modulo gauge equivalence, one can write the Chern-Simons partition function as a path integral
\begin{align}\label{eq:CS path integral}
    \int_{\mathcal{M}} [\mathcal{D}A] e^{-\frac{1}{\hbar}CS(A)}~,
\end{align}
where the Chern-Simons action 
\begin{align}
    CS(A) = \frac{1}{4\pi}\int_{M }\operatorname{Tr}\left(A \wedge \d A + \frac{2}{3} A\wedge A\wedge A \right)\,
\end{align}
is a holomorphic function on $\mathcal{M}$ and $\frac{1}{\hbar}\equiv k \in \mathbb{C} \backslash \{0\}$ is the complexified level of Chern-Simons theory.
The definition of the path integral is over all gauge equivalence classes of connections, however, due to its topological nature, the integral \eqref{eq:CS path integral} localises and the dominant contributions come from neighbourhoods of complex \emph{flat} connections.
These are the classical solutions, or saddles of the Chern-Simons action, satisfying vanishing curvature field strength
\begin{equation}
F_{A_\alpha}= dA_\alpha+ A_\alpha \wedge A_\alpha=0\,,
\end{equation}
and similarly for the anti-holomorphic part.

While 
$\rm SU(2)$ connections on a 3-manifold form a compact moduli space, the same is not true for gauge group $\rm SL(2,\mathbb{C})$ which allows for sequences of connections whose limit is singular (we will discuss them in greater detail in next Section). However, it was conjectured in \cite{gukov2024categorification} that such ``infinite asymptotic ends'' can have a finite Chern-Simons action despite not being classical saddles of that action, and thus can be detectable using the tools of resurgence. Below we will briefly present the resurgent analysis for this theory.

\paragraph{Resurgence in a nutshell.}

Starting with \eqref{eq:CS path integral}, we can write the formal perturbative expansion of the Chern-Simons partition function 
around each of its classical saddles $\alpha$ as
\begin{equation}
Z_\alpha^{\rm pert}[M;\hbar]:= e^{- {CS_{\alpha}}/{\hbar}}\sum_{n=0}^{\infty} \hbar^{(n+c_\alpha)} a_\alpha^{(n)}.
\label{pertseries}
\end{equation}
where $CS_\alpha$ is the Chern Simons action evaluated at the saddle $\alpha$, $c_\alpha$ is a constant and $a_{\alpha}^{(n)}$ denotes the $n$th perturbative coefficient. The goal of resurgence is, roughly speaking, to analyse the collection of perturbative coefficients around each saddle and to unify them into one large non-perturbative picture.
Note that such a series \eqref{pertseries} in variable $\hbar$ is in general divergent. There is a procedure called {\it Borel transform} to convert such divergent series into a  convergent one, provided the growth of $a_\alpha^{(n)}$ for large $n$ is known.  For instance, if  $a_\alpha^{(n)}\sim \Gamma (n+c_\alpha)$, 
the Borel transform of $Z_\alpha^{\rm pert}[M;\hbar]$  is defined as
\begin{equation}
B_\alpha[\xi] = \sum_{n=0}^\infty \frac{a_{\alpha}^{(n)} }{\Gamma(n+c_\alpha) } \, (\xi-CS_{\alpha})^{n+c_\alpha-1}.
\label{borel}
\end{equation}
We can determine the singularities of the function $B_\alpha[\xi]$ in the complex $\xi$ plane, called the \emph{Borel plane}. By virtue of the resurgence, the singular points
$\{\xi_\star\}$ indicate the presence of other saddles  in the theory, since $$\xi_\star=CS_{\beta}-CS_{\alpha},~$$  where $CS_{\beta}$ denotes the value of the  Chern-Simons action evaluated at a saddle $\beta\neq \alpha$. We can recover the perturbative series (\ref{pertseries}) by performing the Borel resummation
\begin{equation}
    Z_\alpha{([M; \hbar])}\; \simeq \; \int_{CS_\alpha}^{\infty} d \xi \; B_\alpha[\xi]\;\ e^{- \xi/\hbar}\,,
\end{equation}
where the contour of integration in the  $\xi$ plane should be such that it avoids all the singularities of the Borel transform. 


\paragraph{Flat connections on $S^3\setminus K$ and $S^3_{p/r}(K)$.} We will now  discuss the moduli space of flat connections on a three-manifold with a single torus boundary, i.e. a complement of a knot $K$ in $S^3$. For a general 3-manifold $M$ and the gauge group $\rm G$, we have the identification
\begin{equation}
\mathcal{M}_{\rm flat}(M)=\operatorname{Hom}(\pi_1(M),\rm G)/\rm G\,,
\end{equation}
where quotient is defined by the conjugation action of $\rm G$ and $\pi_{1}(M)$ denotes the fundamental group corresponding to $M$. 
Since every flat connection on $S^3 \backslash K$ can be restricted to a flat connection on its boundary torus, we can focus on describing the latter.
The moduli space of flat $\mathrm{SL}(2,\mathbb{C})$ connections on the torus $\Sigma = T^2$ is simply 
conjugacy classes of commuting pairs $(x,y)\in\mathbb{C}^* \times \mathbb{C}^*$ corresponding to the eigenvalues of the holonomies around the two fundamental cycles, \textit{meridian} $\mu$ and \textit{longitude} $\lambda$, of that torus. Indeed, since $\pi_1(\partial M)$ is abelian, the images $\rho(\mu)$ and $\rho(\lambda)$ commute, and they can be simultaneously diagonalised.
Due to an action of $\mathbb{Z}_2:(x,y) \rightarrow (x^{-1}, y^{-1})$, we can identify 
\begin{equation}
\mathcal{M}_{\rm flat}(T^2) \simeq \left( \mathbb{C}^* \times \mathbb{C}^* \right) / \mathbb{Z}_2\,.
\end{equation} 
\noindent Note that this space becomes singular at the four points: 
\begin{equation}\label{eq:singular points}
x = \pm 1 ~~{\rm and}~~y = \pm 1~,
\end{equation}
as they are the fixed points under the $\mathbb{Z}_2$ action.

In order to be extendable from the boundary to the bulk of our 3-manifold, flat connections on the boundary torus are required to satisfy knot group relations. Restricting this to the peripheral subgroup, we get the algebraic condition
\begin{equation}\label{eq:A-poly constraint}
    \mathcal{A}_{K}(x,y)=0\,,
\end{equation}
where $\mathcal{A}_{K}(x,y)=(y-1)\mathcal{A}^{\rm irr}_K(x,y)$ denotes the $A$-polynomial of the associated knot $K$ \cite{cooper1994plane,cooper1998representation}. This gives the almost surjective map
\begin{equation}\label{eq:mflat complement}
\mathcal{M}_{\rm flat}(S^3\setminus K)\ \xrightarrow{\ \sigma\ }\ \{(x,y)\ \vert\ (x,y)\in\left( \mathbb{C}^* \times \mathbb{C}^* \right) / \mathbb{Z}_2,\ \mathcal{A}_K(x,y)=0 \}\,.
\end{equation} 
Here by ``almost surjective'' we mean that it is surjective apart from a discrete set which includes \eqref{eq:singular points}. That is, some of the singular points \eqref{eq:singular points} do not correspond to the usual flat connections.\footnote{Apart from $(x,y)=(1,1)$, which corresponds to abelian flat connections.} Note that the resulting set carved out by the $A$-polynomial is not countable. Therefore, in order to count flat connections properly, we need to consider a closed 3-manifold instead. That is, we need to perform a Dehn filling on the torus boundary component $\partial M$.

It is well-known that any closed three-manifold $M$ can be constructed by performing Dehn surgery on some link in $S^3$ \cite{lickorish1962representation}. A particular class of manifolds is then given by surgeries on one-component link, i.e. a knot. Such construction is described by a choice of a knot and a slope $p/r$ where $p$ and $r$ are co-prime integers. Apart from the $A$-polynomial condition \eqref{eq:A-poly constraint}, the holonomy eigenvalues must satisfy the Dehn filling constraint for such a surgery:
\begin{equation}
x^p y^{r} =1\,.
\label{SurgeryCondition}
\end{equation}
As a result, the moduli space of flat connections on a knot surgery manifold $S^3_{p/r}(K)$ admits an almost surjective mapping
\begin{equation}
\mathcal{M}_{\rm flat}(S^3_{p/r}(K))\ \xrightarrow{\ \sigma\ }\ \{(x,y)\ \vert\ (x,y)\in\left( \mathbb{C}^* \times \mathbb{C}^* \right) / \mathbb{Z}_2,\ \mathcal{A}_K(x,y)=0,\ x^py^r=1 \}\,.
\end{equation} 
However, unlike \eqref{eq:mflat complement}, this set is now discrete, which allows us to define a proper count. We will go back to this in Section \ref{sec: counting} and in a similar fashion suggest a way to count flat connections at infinity.

Finally, one can restrict this to irreducible flat connections by considering the irreducible part of the $A$-polynomial $\mathcal A^{\mathrm{irr}}_{K}(x,y)$,
\begin{equation}\label{eq:mflat irr}
\mathcal{M}^{\rm irr}_{\rm flat}(S^3_{p/r}(K))\ \xrightarrow{\ \sigma\ }\ \{(x,y)\ \vert\ (x,y)\in\left( \mathbb{C}^* \times \mathbb{C}^* \right) / \mathbb{Z}_2,\ \mathcal{A}_{K}^{\rm irr}(x,y)=0,\ x^py^r=1 \}\,.
\end{equation} 
The moduli space $\mathcal{M}_{\rm flat}^{\rm irr}$ (more precisely, its compactification which will be introduced in the next Section) and the map $\sigma$ will be of our main interest in this paper.

\paragraph{Numerical Chern-Simons action.}
Lastly, we review the computation of the Chern-Simons action $CS_{\alpha}$ corresponding to an element $\alpha\in \mathcal{M}_{\rm flat}^{\rm irr}(M)$ for $M= S^3_{p/r}(K)$. More precisely, these will determine the numerical (lifted) Chern-Simons invariants corresponding to flat connections lifted to the universal cover $\widetilde{M}$ of $M$.
We will closely follow the steps from \cite{kirk1990chern,Costin:2023kla}:
\begin{itemize}
\item Taking advantage of the map $\sigma$ and \eqref{eq:mflat irr}, we determine the set
    \begin{align}
\left(\mathcal{A}_K^{\mathrm{irr}}(x, y) = 0 \right) \cap \left( 
x^{p} y^{r} = 1\right)~.   \label{ApolySubtoSurgery} 
\end{align} 
\item Let $\Delta_K(x)$ denote the Alexander polynomial of $K$.
For each solution $(x_{\alpha}, y_{\alpha})$ of (\ref{ApolySubtoSurgery}) and for every $x_0\in \Delta_0 := \{x\ | \ \Delta_K(x^2)=0\}$, we construct an integral over a  path $\gamma_{\alpha} \subset \{(x,y)\ |\ \mathcal{A}_K^{\rm irr}(x,y)=0\}$ from $(x_0,y_0=1)$ to $(x_{\alpha},y_{\alpha})$  which gives
\begin{align}
    2\pi^2 CS_\alpha[M] &=\int_{\gamma_\alpha}\frac{\log(y)}{x}dx+\frac{ip}{2}\log\left(x_{\alpha}\right)^2+\frac{jr}{2}\log\left(y_{\alpha}\right)^2-ir\log(x_{\alpha})\log(y_{\alpha})\label{CSaction}
\end{align}
where the pair of integers $i,j$ satisfy $pj-ri=1$.

\end{itemize}
 These critical values of Chern-Simons action $\{CS_{\alpha}[M]\}$ for all the flat connections associated to the three-manifold $M$ can be compared for some three-manifolds which can be given a  plumbed graph description.
 We will also outline the procedure for such plumbed three-manifolds in Appendix \ref{plumbed methd}.



\section{Flat connections at infinity} \label{sec: flat}
Recall from Section \ref{sec:prereq} that some singular points \eqref{eq:singular points} do not correspond to
the usual flat connections. Instead, we conjecture that they correspond to flat connections at infinity.
Thus, by considering a suitable compactification of $\mathcal{M}_{\rm flat}$,
\begin{equation}
\bar{\mathcal{M}}_{\rm flat}(M):=\operatorname{Hom}(\pi_1(M),\rm G)/\rm G \ \cup\ \{\text{flat conn. at $\infty$}\}\,,
\end{equation}
we may write
\begin{equation}\label{eq:mbar flat 2}
\bar{\mathcal{M}}_{\rm flat}(S^3\setminus K)\ \xrightarrow{\ \bar{\sigma} \ }\ \{(x,y)\ \vert\ (x,y)\in\left( \mathbb{C}^* \times \mathbb{C}^* \right) / \mathbb{Z}_2,\ \mathcal{A}_K(x,y)=0 \}\,,
\end{equation} 
where $\bar{\sigma}$ is now surjective\footnote{Here we stress that $\sigma$ is surjective for all non-singular points, but fails to be such for singular points. In comparison, $\overline{\sigma}$ is surjective on the full set of intersection points, including singular ones.} (and similarly for knot surgery manifold $M=S^3_{p/r}(K)$). By analogy to \eqref{eq:mflat irr}, we can also consider a suitable compactification of $\bar{\mathcal{M}}_{\rm flat}^{\rm irr}(M)$ which also yields a surjective map $\bar{\sigma}$.

The driving observation for us is that each of the four points \eqref{eq:singular points} in the image of $\bar{\mathcal{M}}_{\rm flat}^{\rm irr}$ correspond to a new branch of flat connection implying that the holonomies can be continuously deformed into upper-triangular matrices with unipotent structures \cite{Chun:2019mal}. Specifically, the holonomies $X$ and $Y$ around the meridian and longitude cycles of the torus can take the form
\begin{equation}
X =
\begin{pmatrix}
\pm 1 & u_1 \\
0 & \pm 1
\end{pmatrix},
\qquad
Y =
\begin{pmatrix}
\pm 1 & u_2 \\
0 & \pm 1
\end{pmatrix},
\end{equation}
where $u_1$ and $u_2$ are complex parameters that describe the deformation away from the diagonal holonomies.\footnote{The explicit form of these parameters for a particular knot can be deduced from the knot group relations.} We will focus on one such deformation arising from one of those fixed points. 

Using a developing map \cite{culler1983varieties, cooper1998representation} one can construct flat connections with the deformed holonomies $X$ and $Y$ (see also Section \ref{3_1 example}). Since the moduli space of complex flat connections on a general 3-manifold is non-compact, there can exist families $\{A_
{1/\epsilon}\}$ of (generically) non-flat connections parametrised by $\epsilon$ which can be either discrete or continuous, such that some components tend to infinity as the parameter $\epsilon \rightarrow 0$, but the curvature tends to zero.
Following \cite{gukov2024categorification}, we refer to them as \emph{flat connections at infinity} and denote $A_{\infty}$.
One can start by constructing a family of such connections for knot complements, where the meridian and longitude holonomies are given by a $2 \times 2$ unipotent matrix of the form:
\begin{equation}
X \; \equiv \rho({\mu})= \;
\begin{pmatrix}
x & \epsilon \\
0 & x^{-1}
\end{pmatrix},\qquad 
Y \equiv \rho({\lambda})=
\begin{pmatrix}
y & * \\
0 & y^{-1}
\end{pmatrix}.
\label{y1twist}
\end{equation}
For instance, one can choose $x=1$ and $y=-1$, which corresponds to one of the singular points \eqref{eq:singular points}. The only condition for the off-diagonal element ``*'' is its regularity as $\epsilon\rightarrow 0$. As we will also see in the examples, such flat connections $A_{1/\epsilon}$ for a finite $\epsilon\neq 0$ cannot be extended from knot complement to a surgery manifold. However, in the limit $\epsilon\to 0$, the curvature vanishes and we indeed obtain a new kind of flat connection on a closed 3-manifold.

Next, we will outline our conjectural approach to counting all flat connections, including those at infinity, for a class of closed 3-manifolds.

\subsection{Counting points in \texorpdfstring{$\bar{\mathcal{M}}_{\rm flat}^{\rm irr}$}{moflat}}\label{sec: counting}

The main goal of our work is to find and tabulate such flat connections at infinity for complements of torus and double twist knots, as well as $\pm1/r$-surgeries on them. To this end, we rely on two different ideas -- the first one is more general but computationally expensive, while the other one is much cheaper but applies only for a class of integer homology 3-spheres. In the end, the two approaches can be combined and tested for a large class of knot surgeries; in Section \ref{sec: case studies} we study them in numerous examples.

\paragraph{The first approach.}

We begin with the fundamental group $\pi_1$ of the knot complement $S^3 \setminus K$.
The Wirtinger presentation of $\pi_1(S^3 \setminus K)$ for a general knot $K$ is constructed by assigning generators to the arcs in the planar diagram of the knot -- such generators satisfy relations based on the crossings. This results in  two independent generators of the fundamental group, denoted $a$ and $b$. We identify generator $a$ with the meridian of the knot. The longitude is obtained as a function of the generators $a$ and $b$. We then consider the homomorphism $\rho: \pi_{1}(S^{3}\setminus K)\rightarrow SL(2,\mathbb{C})$, under which the Wirtinger generators $a$ and $b$ admit the following $2\times 2$ matrix form with complex entries (analogous form was considered in \cite{gukov2024categorification})
\begin{equation}
\rho(b) = 
\begin{pmatrix}
\ast & \ast \\
\frac{C}{\epsilon} & \ast
\end{pmatrix},
\qquad
\rho(a) =
\begin{pmatrix}
1 & \epsilon \\
0 & 1
\end{pmatrix},
\label{generators}
\end{equation}
where $C\in \mathbb{C}$ is a constant, and each “$\ast$” represents the function of $\epsilon$ that remains regular in the limit $\epsilon \to 0$. 

The steps involved in  finding $A_{\infty}$  can be summarized as follows:
\begin{enumerate}
    \item From the Wirtinger presentation, we choose explicit matrix forms for the meridian, longitude, and the generators of the fundamental group (\ref{y1twist}, \ref{generators}).
    
    \item Then, we solve the fundamental group relation by assuming that form of the generators, as well as by taking the eigenvalues $x=1$ and $y=-1$.
    
    \item Finally, we count the total number of nonzero values of the parameter $C$ that arise from the previous step; in order to confirm that the resulting flat connection extends to the surgery manifold, we take $\epsilon \rightarrow 0$ and verify the surgery relation $X^pY^r=1$.
\end{enumerate}
These nonzero values of $C$ define one possible way to count flat connections at infinity. 

\paragraph{The second approach.}
Our starting point is the following observation, already mentioned in Section \ref{sec:prereq}. The moduli space of irreducible complex flat connections on a general knot surgery manifold can be mapped almost surjectively onto the set of pairs of complex non-zero numbers $(x,y)$. Furthermore, this set of pairs must satisfy the $A$-polynomial relation plus the surgery (Dehn filling) relation.
Following the above discussion, we assume that singular points \eqref{eq:singular points} which are obstructions to a surjection, in fact correspond to flat connections at infinity. We further conjecture that in order to properly count asymptotic ends of $\bar{\mathcal{M}}_{\rm flat}^{\rm irr}$ corresponding to such points, it is sufficient to determine the overall degree
of the map $\bar{\sigma}$ in
\begin{equation}\label{eq:mirr surjection}
\bar{\mathcal{M}}_{\rm flat}(S^3_{p/r}(K))\ \xrightarrow{\ \bar{\sigma} \ }\ \{(x,y)\ \vert\ (x,y)\in\left( \mathbb{C}^* \times \mathbb{C}^* \right) / \mathbb{Z}_2,\ \mathcal{A}_K(x,y)=0,\ x^py^r=1 \}\,.
\end{equation} 

In other words, our idea is to scan the images of $\sigma$ for every non-singular point in $A(x,y) = 0 \cap {x^p}{y^r} = 1$, and count how many usual flat connections correspond to each such point. Then, since the map $\sigma$ is surjective, we expect $\overline{\sigma}$ to be surjective too. By including singular points, we can see that each one of them (in our case studies, we are focusing on $(x, y) = (1, -1)$) also corresponds to some number of connections, and this should be the same number as above, computed for the usual flat connections. However, singular points cannot correspond to usual flat connections, as seen from comparing their Chern-Simons invariants -- instead, we expect them to correspond to flat connections at infinity.

Suppose that $M$ is a Brieskorn sphere coming from a rational surgery on a torus knot $T(s,t)$, and let $n$ be the total number of points in the codomain of $\bar{\sigma}$ in \eqref{eq:mirr surjection} that includes singular points. Denote $\Delta_0^+ := \{x\ | \ \Delta_K(x^2)=0,\ 0\leq \arg x \leq \pi/2 \}$, and let $m=|\Delta_0^+|$.
Fix one point in the codomain of $\bar{\sigma}$ as a base point $(x_\alpha,y_\alpha)$, and compute $m$ integrals \eqref{CSaction} over a straight path from $(x_\alpha,y_\alpha)$ to $(x_0,y(x_0))$ for every $x_0\in \Delta_0^+$ (note that for $T(s,t)$ the integration domain is simply connected). As a result, we find $m$ numerical Chern-Simons values corresponding to this base point. Some of these values may correspond to the same Chern-Simons invariant modulo 1; however, each of them would produce an invariant of a different flat connection. By applying this idea to $M=S^3_{\pm1/r}(T(s,t))$ for even values of $r$, we find experimentally the degree of $\bar{\sigma}$,
\begin{equation}
\operatorname{deg}\bar{\sigma} = \frac{(s-1)(t-1)}{2}\,. 
\end{equation}

If we take into account singular points \eqref{eq:singular points} and the surgery condition, we can write the conjectural refinement of the $\rm{SL}(2,\mathbb{C})$ Casson invariant (compare with \cite{boden2006sl}) which takes into account flat connections at infinity:\footnote{It would be very interesting to extend this conjecture beyond torus knot surgeries -- we leave this for future research.}
\begin{equation}
    \overline{\lambda_{\rm SL(2,\mathbb{C})}}(\Sigma(s,t,rst\pm 1)) := \frac{(s-1)(t-1)((rst\pm 1)-1)}{4} + \frac{(s-1)(t-1)}{2}\,.
\end{equation}

\paragraph{} In the remaining part of the paper, we will test the above approaches for a class of torus and double twist knots (for the latter, though, the integration domain is not simply connected and a more precise refinement is required -- we thus resort to the first approach only).

\section{Case studies}\label{sec: case studies}

Our goal in this Section is to detail the computation of flat connections at infinity for some torus, twist as well as double twist knot surgeries (in all examples knots are considered in $S^3$). In other words, we are able to count points in $\bar{\mathcal{M}}_{\rm flat}^{\rm irr}$ using both approaches outlined in Section \ref{sec: counting}, and confirm their agreement. These results can be thought of as a first step towards a broader, more systematic understanding of flat connections at infinity for a larger class of 3-manifolds.

\subsection{Detailed example: rational surgery on the trefoil knot}\label{3_1 example}

\begin{figure}
    \centering
    \includegraphics[width=0.55\linewidth]{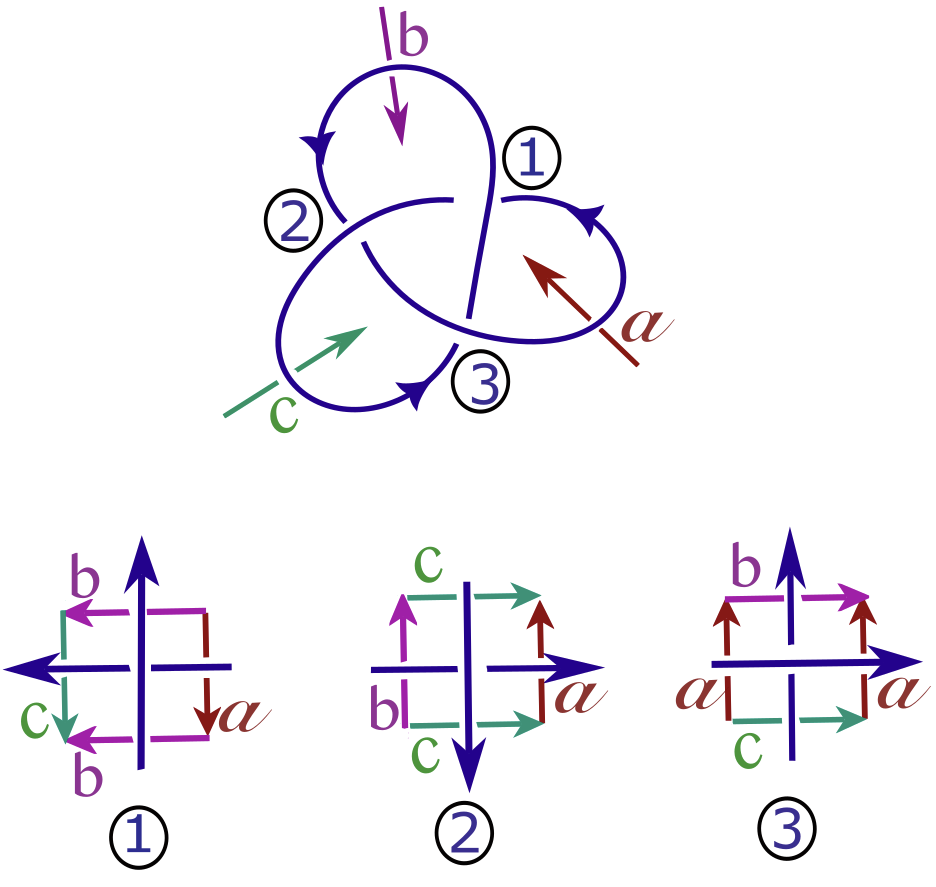}
    \caption{Wirtinger presentation of the trefoil knot complement $S^3\setminus T(2,3)$}
\label{WirRepRHTrefoil}
\end{figure}

We begin this example with computing the fundamental group of the complement of the trefoil knot $T(2,3)$ using the Wirtinger presentation -- that is, we follow the first approach from Section \ref{sec: counting}. For the  trefoil knot we get the following relations from the loops drawn at its three crossings (Figure \ref{WirRepRHTrefoil})
\begin{equation}
ab=bc, \,\ ~~~bc=ca, ~~~ab=ca\,.
\label{trfoil relations}
\end{equation}

\noindent We express $c$ in terms of $a$ and $b$ from \eqref{trfoil relations} and further simplify the above relations to obtain the fundamental group corresponding to the complement of trefoil knot:
\begin{align}
    \pi_{1}(S^3\setminus T(2,3))&=\langle  a,b;\ bab=aba\rangle\,.
    \label{fundamenta group trefoil}
\end{align}
 For the  trefoil knot,   the meridian $\mu$ and the longitude $\lambda$ are given in terms of Wirtinger generators $a,~b$  as
 \begin{equation}
  \mu= a~;~  \lambda=a^{-3}bac=a^{-3}bab^{-1}ab\,.\label{longitude trefoil}
\end{equation}
The factor $a^{-3}$ is included in the expression of $\lambda$ so that the linking number of the longitude with the trefoil vanishes.
As a next step, we take the $\rm SL(2,\mathbb{C})$ matrix representation of the generators as given in \eqref{generators} to obtain the matrix form for the meridian and longitude 
\begin{equation}\label{eq:rho constraints trefoil}
X=\rho(a),~~Y=\rho(a)^{-3}\rho(b)\rho(a)\rho(b)^{-1}\rho(a)\rho(b)\,.
\end{equation}
Taking the ansatz (\ref{y1twist}) and \eqref{generators}, setting $x=1$ and $y=-1$ and using the fundamental group relation 
\begin{align}
\rho(b)\rho(a)\rho(b)=\rho(a)\rho(b)\rho(a),
\end{align}
we get unique solution with $C=-1$:
\begin{equation}\label{trefoil fc infty}
    X =
    \begin{pmatrix}
        1 & \epsilon \\
        0 & 1
    \end{pmatrix}
    \,,\qquad Y = 
    \begin{pmatrix}
        -1 & 6\epsilon \\
        0  & -1
    \end{pmatrix}
    \,,
    \qquad
    \rho(b) = 
    \begin{pmatrix}
        \beta & \epsilon(\beta-1)^2 \\
        -\frac{1}{\epsilon} & 2-\beta \\
    \end{pmatrix}
    \,,
\end{equation}
where $\beta\in \mathbb{C}$.
As a result, after performing $\pm1/r$-surgery, we obtain a family of non-abelian connections parametrised by $1/\epsilon$. Note that the surgery constraint
\begin{equation}
    X Y^{\pm r} = 1\,
\end{equation}
is satisfied in the $\epsilon\to 0$ limit, only when $r$ is an even integer. In other words, we have a family of non-flat connections on a surgery manifold, which become flat in the above limit, for even surgeries.

Note that $\pm 1/r$-surgery on a torus knot produces a Brieskorn homology sphere, which is a Seifert fibred 3-manifold with three singular fibres and has a plumbing graph description as a ``3-leg star graph'' \cite{Cheng:2022rqr}. For example, $-1/r$ surgery gives rise to the following class of Brieskorn sphere
\begin{equation}\label{eq:brieskorn sphere}
 S^3_{-1/r} (T(s,\pm t)) = \Sigma(s, t, rst \pm 1)\,. 
\end{equation}
The resurgent structure of the Chern-Simons path integral is well-known in this case \cite{Gukov:2016njj,AM22}. In particular, it is established that only ordinary, finite flat connection appear in the resurgent formulation of \eqref{eq:CS path integral} (in other words, flat connections at infinity do not appear as singularities on the Borel plane).
On the other hand, $+1/r$-surgery produces the reverse-oriented Brieskorn sphere:
\begin{equation}\label{eq:brieskorn sphere inversed}
 S^3_{1/r} (T(s,\pm t)) = \overline{\Sigma(s, t, rst \mp 1)}\,. 
\end{equation}
 This class of examples is much less explored from the resurgence perspective (notable prior works include \cite{Costin:2023kla,Adams:2025aad}).  It is possible that the path integral \eqref{eq:CS path integral} may involve contributions of flat connections at infinity, apart from the usual flat connections. One such potential example is $+1/2$-surgery on $T(2,-3)$ which gives Brieskorn sphere $\overline{\Sigma(2,3,13)}$. In \cite{Harichurn:2025suf}, the authors assumed the mock modular property of homological blocks, i.e. $\widehat{Z}$ invariants, and compared the modular and resurgent decompositions of such blocks for $\overline{\Sigma(2,3,13)}$ in particular. It turned out that some terms in such decomposition do not correspond to the usual flat connections -- possibly, they indicate the presence of flat connections at infinity. The above said deserve a further thorough study which we leave for future; below we will focus on the simpler analysis of a flat connection at infinity for $\overline{\Sigma(2,3,13)}$, following the second approach from Section \ref{sec: counting}.
 
\begin{figure}[h!]
\centering
\includegraphics[width=0.65\textwidth]{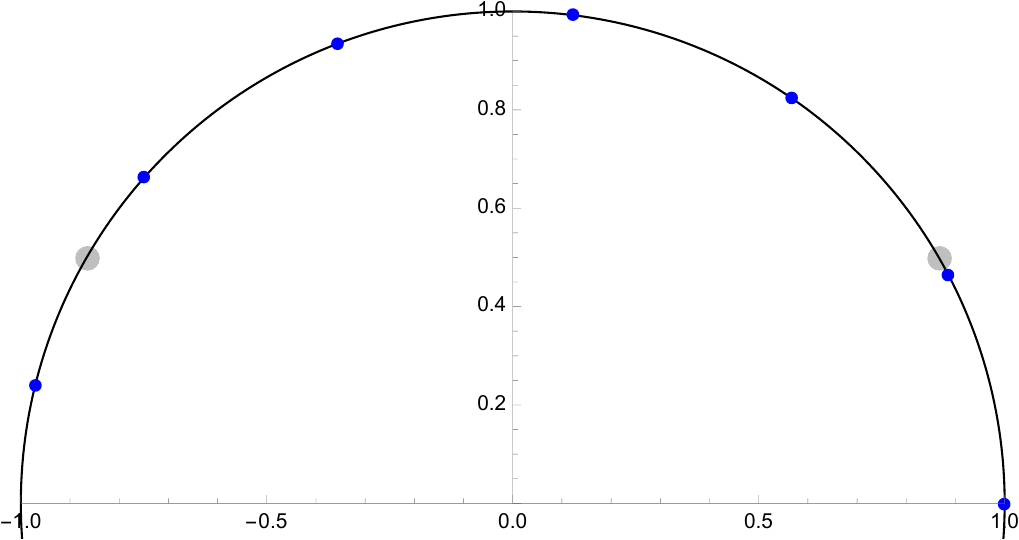}
\caption{Plot of $x$-values: $x_{\alpha}$ (blue dots) and zeros of the Alexander polynomial $\{x\,\vert\, \Delta_{3_1}(x^2)=0\}$ (gray dots) for $S^3_{1/2}(T(2,-3)) = \overline{\Sigma(2,3,13)}$}
\label{CS path trefoil2-3}
\end{figure}

  Recall that the $A$-polynomial of $T(2,-3)$ is
\begin{equation}
    \mathcal{A}_{T(2,-3)}(x, y) = (y-1)(x^6 + y)\,.
    \label{Apoly trefoil}
\end{equation}
 Computing the intersection of $\mathcal{A}^{\rm irr}_{T(2,-3)}(x,y)=0$ with $xy^{2}=1$, we get 6 points as tabulated in Table \ref{values of x2-3} including a singular point corresponding to $x_{\alpha^*}=1$. Their images on the $x$-plane are shown as blue dots in Figure \ref{CS path trefoil2-3} (due to the Weyl symmetry, we consider only the upper half plane).

\begin{table}[t!]
\centering
\begin{tabular}{||c|c||}
\hline
\textbf{Index} & \textbf{Value of $x_{\alpha}$} \\
\hline\hline
1 & $-0.970942+0.239316 i$ \\
\hline
2 & $-0.748511+0.663123 i$ \\
\hline
3 & $-0.354605+0.935016 i$ \\
\hline
4 & $0.120537\, +0.992709 i$ \\
\hline
5 & $0.568065\, +0.822984 i$ \\
\hline
6 & $0.885456\, +0.464723 i$ \\
\hline
7 & $1$ \\
\hline
\end{tabular}
\caption{$x_{\alpha}$ values obtained by solving $\mathcal{A}^{\rm irr}_{T(2,-3)}(x,\pm x^{-1/2}) = 0$ }
\label{values of x2-3}
\end{table} 

The zeroes of the Alexander polynomial 
$\Delta_{3_1}(x^2) = x^2 + x^{-2} - 1$
are 
$$
\Delta_0 = \left\{ \pm e^{\pi i/6},\ \pm e^{5\pi i/6} \right\}\,.$$
Zeroes from the upper half of complex $x$--plane $\Delta_0^+\subset \Delta_0$ are shown as gray dots in Figure~\ref{CS path trefoil2-3}.
To compute the Chern-Simons invariants, we 
choose the initial point as $x_0\in\Delta_0^+ =e^{\pi i/6}$.  In order to compute the Chern-Simons action, we follow a  path from every intersection point $x_{\alpha}$ including $x_{\alpha^*} = 1$ (refer to Table \ref{values of x2-3}) to $x_0 = e^{\pi i/6}$ and evaluate the integral~\eqref{CSaction}.
Numerical values of the Chern-Simons action for  all the flat connections turns out to be (corresponding to the blue points in Figure \ref{CS path trefoil2-3} in the clockwise direction from $x_1$ to $x_7$):
\begin{equation}
    \{CS_\alpha\}=\{
 \frac{7^2}{312},
 \frac{5^2}{312},
 \frac{17^2}{312},
 \frac{23^2}{312},
 \frac{11^2}{312},
 \frac{1^2}{312}, \underbrace{\frac{13^2}{312}}_{\alpha^*}
 \}~~\mathrm{mod}~1
 \label{num cs trefoil}
\end{equation}
In this case the Casson invariant $\lambda_{\rm SL(2,\mathbb{C})}(\overline{\Sigma(2,3,13)})=6$, and we determine $\deg \bar\sigma=1$, meaning that every non-singular intersection point (of which there are six in Figure \ref{CS path trefoil2-3}) is the image of a unique element of $\mathcal{M}_{\rm flat}^{\rm irr}$. 
We also find that the Chern-Simons action associated to the singular point $(x,y)=(1,-1)$ is $CS_{\alpha^*} = \frac{13^2}{312}\mod 1$. According to our proposal in Section \ref{sec: counting}, this corresponds to the unique flat connection at infinity given by \eqref{trefoil fc infty}.

A similar calculation can be done for the $-1/2$ surgery on the torus knot $T(2,3)$ which yields   Brieskorn sphere $\Sigma(2,3,13)$ -- a negative definite plumbed 3-manifold.
Note that the irreducible component of the $A$-polynomial is $\mathcal{A}^{\text{irr}}_{T(2,3)}(x, y) = 1+ x^6 y,$
whose intersection with $xy^{-2}=1$ gives the same picture as in Figure \ref{CS path trefoil2-3} above. Similarly, the Chern-Simons invariants are
\begin{equation}
    \{CS_\alpha\}=\{
 -\frac{7^2}{312},
 -\frac{5^2}{312},
 -\frac{17^2}{312},
 -\frac{23^2}{312},
 -\frac{11^2}{312},
 -\frac{1^2}{312}, \underbrace{-\frac{13^2}{312}}_{\alpha^*}
 \}~~\mathrm{mod}~1
 \label{num cs trefoilbr}
\end{equation}
Note the above equation for $\Sigma(2,3,13)$ is related to Chern-Simons invariants of reverse oriented Brieskorn sphere
\eqref{num cs trefoil} by a negative sign. Recall that the flat connection at infinity $CS_{\alpha^\star}=-\frac{13^2}{312} ~\mod ~1$, corresponding to negative definite plumbed manifold is not seen as a singularity on the Borel plane \cite{AM22}. Even though it is not interesting from the resurgent analysis perspective, we can explore whether they are present or absent from the negative plumbed graph method discussed in Appendix \ref{plumbed methd}.


Brieskorn spheres of the type \eqref{eq:brieskorn sphere}, admit plumbed graph 
presentations with negative definite plumbing matrix, as described in~\cite{Gukov:2019mnk, Harichurn:2023akp}.
Following the tools elaborated in Appendix \ref{plumbed methd}, we will now compute the Chern-Simons action for the flat connections from the plumbing matrix.
The plumbing graph for $\Sigma(2, 3, 13)$ is shown in Figure~\ref{fig:(2,3,13)} and has only one high-valency vertex.\footnote{Here by high valency we mean valency $\geq 3$, using the same terminology as in \cite{gukov2024categorification}.}

\begin{figure}[h!]
    \centering
\includegraphics[width=0.5\linewidth]{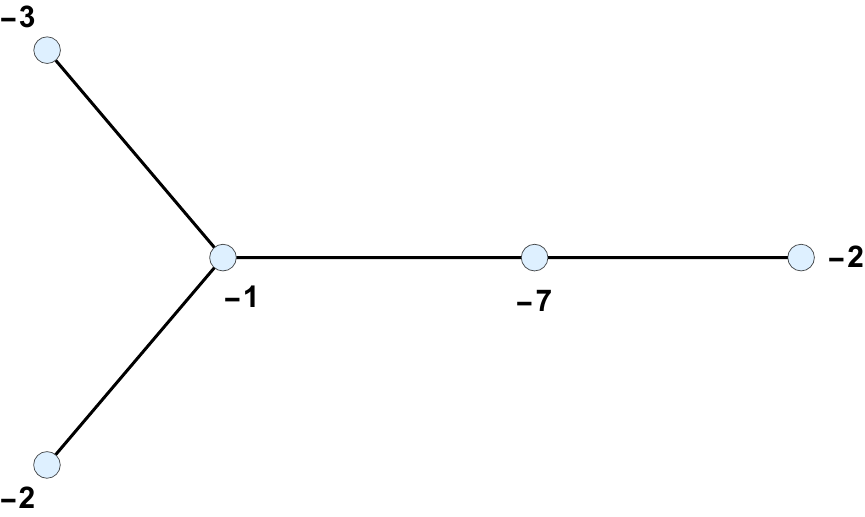}
    \caption{Plumbing graph representation for $\Sigma(2, 3, 13)$}
    \label{fig:(2,3,13)}
\end{figure}

\noindent The adjacency matrix for this 3-manifold 
\begin{equation}
\mathbf{B}^{(\Gamma)} =
\left(
\begin{array}{r|rrrr}
-1 & 1 & 1 & 1 & 0 \\
\hline
1 & -2 & 0 & 0 & 0 \\
1 &  0 & -3 & 0 & 0 \\
1 &  0 &  0 & -7 & 1 \\
0 &  0 &  0 & 1 & -2 \\
\end{array}
\right)~,
\label{bmatrix}
\end{equation}
whose determinant is 1.
Inverse of the matrix $\mathbf{B}$ is:
\begin{equation}
\mathbf{B^{(\Gamma)}}^{-1} =
\left(
\begin{array}{r|rrrr}
-78 & -39 & -26 & -12 & -6 \\
\hline
-39 & -20 & -13 & -6  & -3 \\
-26 & -13 & -9  & -4  & -2 \\
-12 & -6  & -4  & -2  & -1 \\
-6  & -3  & -2  & -1  & -1 \\
\end{array}
\right).
\label{amatrix}
\end{equation}
Following the eqns. \eqref{ContFracMethod}, \eqref{eq:R1_function}, we will get  {only the six finite flat connections}.  

In order to obtain the additional flat-connection at infinity, we propose the following modification in the computation of $R(v_1) \rightarrow 
\tilde R(v_1)$ in eqn. \eqref{eq:R1_function}. Basically,  $P_a$ \eqref{ContFracMethod} for the low-valent vertex that is connected to the high-valent vertex through the bivalent vertex with framing $-7$ is evaluated ignoring the bi-valent vertex.  This leads to 
\begin{equation}
    {\tilde R}({v}_1)=\frac{\sin{\frac{\pi v_1}{2}}\sin{\frac{\pi v_1}{3}}\sin{\frac{\pi v_1}{2}}}{\sin{\pi v_1}}.
     \label{rvplumed1}
\end{equation}
Interestingly, this proposed modification  gives the Chern-Simons action for all the  flat connections including the flat-connection at infinity $CS_{\alpha^*}=\frac{11}{24}$,  obtained using the path-method. In fact, we validated the above proposal for other torus knots-namely, $T(2,5)$, $T(3,4)$ and $T(3,5)$.

For completeness, we will obtain the matrix form of $A_{\alpha^*}$ corresponding to Chern-Simons action $CS_{\alpha^*}(\Sigma(2,3,13) \equiv S^3_{-1/2}(T(2,3))$ using the developing map discussed in Appendix~\ref{gauge connection}.  Using the meridian and longitude holonomies (\ref{y1twist}), the developing map $D({x}_1, {y}_1)$ for trefoil is given by
\begin{equation}
D({x_1}, {y_1}) =
\begin{pmatrix}
\exp\left( \frac{i}{2}( 2x_1 + y_1) \right) & -\frac{(\epsilon x_1 + b_1 y_1)}{2\pi} \exp\left( \frac{i}{2}(2x_1 -y_1)\right) \\
0 & \exp\left( -\frac{i}{2}(2 x_1 +y_1) \right)
\end{pmatrix}
    \label{dev map}
\end{equation}
where $\epsilon$ and $b_1$ are the upper triangular element of the holonomy $X$ and $Y$ respectively. The value of  $b_1$ for $-1/2$ surgery of trefoil is 
\begin{equation}
b_1=   6 \epsilon~.
\end{equation}
In terms of the developing map, the connection one-form $A_{\alpha*} = -dD \cdot D^{-1}$ is:
\begin{equation}
    A_{\alpha^*} = \frac{1}{2\pi}
\begin{pmatrix}
- i  \, d x_1 -\frac{i}{2} \, d y_1 & ~~~\left( \frac{\epsilon}{2\pi} \, d x_1 + \frac{b_1}{2\pi} \, dy_1 \right) \exp(i(2x_1 +y_1)) \\
0 &  i  \, d x_1 + \frac{i }{2} \, d y_1
\end{pmatrix}.
\label{explicit gauge connection}
\end{equation}
The  presence of the off-diagonal $\epsilon$ dependent elements 
allows for an equivalent gauge connection:
$A_{\alpha^*}\rightarrow ~\tilde{A}_{\alpha^*}$, under a constant large gauge transformation as discussed in Appendix \ref{gauge connection}. For a choice of $\omega={1/\epsilon}$ , some matrix elements of $\tilde A_{\alpha^*}$ becomes singular in the limit $\epsilon \rightarrow 0$.
Therefore, the gauge equivalent class $\{A_{\alpha^*}\}$ is referred to as $A_{1/\epsilon}$. In the limit $\epsilon \rightarrow 0$, we indeed find that the curvature two-form $F_{\alpha^*} = d A_{\alpha^*} + A_{\alpha^*} \wedge A_{\alpha^*}$ vanishes.
This confirms that the $A_{\alpha^*}$ is the flat connection at infinity.

Having finished the detailed analysis for $-1/2$-surgery on the trefoil knot, in the remainder we will address small rational surgeries for other torus, twist and double twist knots. Most importantly, we will confirm the existence of flat connections at infinity in those examples as well.

\subsection{\texorpdfstring{$T(s,t)$}{Tst}~ torus knots}

The fundamental group $\pi_1$ of the $T{(s,t)}$ torus knot, where $s ,t$ are  co-prime, is given by the relation \cite{clay2013left}:
\begin{equation}\label{eq:torus knot group}
 \pi_1(S^3 \setminus T{(s,t)}) = \langle g, h \,|\, g^s = h^t \rangle,  
\end{equation}
where $g$ and $h$ are the two generators of the fundamental group. The explicit form of $g$ and $h$, starting from the Wirtinger presentation generators $a,b,\ldots$ turns out to be
$$g=aba~;~ {\rm and} ~h=ab~.$$ 
The meridian $(\mu)$ and longitude $(\lambda)$, expressed in terms of the generators $g$ and $h$, are given by:
\begin{equation}
    \mu\equiv a=h^{j} g^{i}, \,\ ~~~~~ \lambda=\mu^{-s t} g^s ,
\end{equation}
where the integers $i$ and $j$ satisfy $s j+ t i=1$.
 The irreducible part of the $A$-polynomial of $T(s,t)$ torus knot is 
 \begin{equation}
 \mathcal{A}_{T(s,t)}^{\rm irr}(x,y)= 1 + x^{(s t)} y.
 \end{equation}
 For the generators $g,h \in SL(2,\mathbb{C})$ group, we take the following matrix form:
 \begin{align}\label{g_and_h}
    g=\begin{pmatrix}
        x_1&x_2\\
        x_3&x_4
\end{pmatrix},~~h=\begin{pmatrix}
        \star &\star\\
        \frac{C}{\epsilon}&\star
    \end{pmatrix}~,
\end{align}
 where $x_1,\dots,x_4$ are some variables, and  $\star$ denotes the terms regular in $\epsilon$. Following the first approach described in Section \ref{sec: counting}, we incorporate these generators into torus knot group relation \eqref{eq:torus knot group} and find $m$ non-trivial values of $C$ for $T(2,2m+1)$  torus knot complement. These flat connections at infinity $\alpha^*$ can be further extended under $\pm1/r$ rational surgery when $r$ is even (much like in the previous example of the trefoil knot).
In Table \ref{table:TorusKnot_m2} we present the data of $C=\{C_i\}$ 
  for $T(2,2m+1)$ torus knots with $m\leq 7$. Interestingly, for every fixed $m$ we find the polynomial constructed from the set: $\prod_i (C-C_i)$ is the Conway polynomial of $T(2,2m+1)$, a well-known knot invariant.\footnote{This connection seems surprising for us, and certainly deserves further attention.}

  Similarly, the second approach gives the same estimate on the number of flat connections at infinity -- the degree of the map $\overline \sigma$ in this case equals $(2-1)(2m+1-1)/2=m$.
  Recall that the $-1/r$-rational surgery on this torus knot gives Brieskorn sphere $\Sigma[(2,2m+1,2r(2m+1)+1]$ which also has a plumbed graph description (for $m=1$ see Figure \ref{fig:(2,3,13)}). Hence, our result yields the estimate
  $$|\bar{\mathcal{M}}_{\rm flat}^{\rm irr}(\Sigma(2,2m+1,2r(2m+1)+ 1))|=\frac{4mr(2m+1)}{4}+m\,.$$
  We further validated the Chern-Simons action corresponding to the  finite flat connections and these additional black swans by (i) the path method (ii) the plumbed graph method for low values of $m$ where 
  we had to implement our proposal about modified $P_a$ computation.
  Particularly, we ignore the bi-valent vertex whose framing number magnitude ($|f|$) is highest to reproduce all the flat-connections.

  \begin{table}[htbp]
\centering
\begin{adjustbox}{width=\textwidth}

\begin{tabular}{||c|>{\raggedright\arraybackslash}p{0.25\textwidth}|>{\raggedright\arraybackslash}p{0.45\textwidth}|c||}
\hline 
Knot $T(s,t)$ & $C$ Polynomial  & Zeros of $C$ polynomial & \# $\alpha^*$ \\ \hline\hline
$T_{(2,3)}$ & 
$C + 1$ & 
$-1$ & 1 \\ \hline

$T_{(2,5)}$ & 
$1 + 3C + C^2$ & 
$\frac{1}{2} (-3 - \sqrt{5}), \frac{1}{2} (-3 + \sqrt{5})$ & 2 \\ \hline
$T_{(2,7)}$ & 
$1 + 6C + 5C^2 + C^3$ & 
$-3.2469796, -1.5549581, -0.1980623$ & 3 \\ \hline
$T_{(2,9)}$ & 
$1 + 10C + 15C^2 + 7C^3 + C^4$ & 
$-3.5320889, -2.3472964, -1.0000000,\newline -0.1206148$ & 4 \\ \hline
$T_{(2,11)}$ & 
$1 + 15C + 35C^2 + 28C^3 + 9C^4 + C^5$ & 
$-3.6825071, -2.8308300, -1.7153703,\newline  -0.6902785,-0.0810141$ & 5 \\ \hline
$T_{(2,13)}$ & 
$1 + 21C + 70C^2 + 84C^3 + 45C^4 + 11C^5 + C^6$ & 
$-3.7709121, -3.1361295, -2.2410734,\newline -1.2907902, -0.5029785, -0.0581164$ & 6 \\ \hline
$T_{(2,15)}$ & 
$-1 - 4C + 6C^2 + 10C^3 - 5C^4 - 6C^5 + C^6 + C^7$ & 
$-1.0000000, -2.61803,-0.381966\newline -3.82709, -3.33826, -1.79094,\newline -0.0437048$ & 7 \\ \hline
\end{tabular}
\end{adjustbox}
\caption{Flat connections at infinity for torus knots $T(2,2m+1)$ under rational Dehn surgery}
\label{table:TorusKnot_m2}
\end{table}

\begin{figure}[h!]
    \centering  \includegraphics[width=0.6\linewidth]{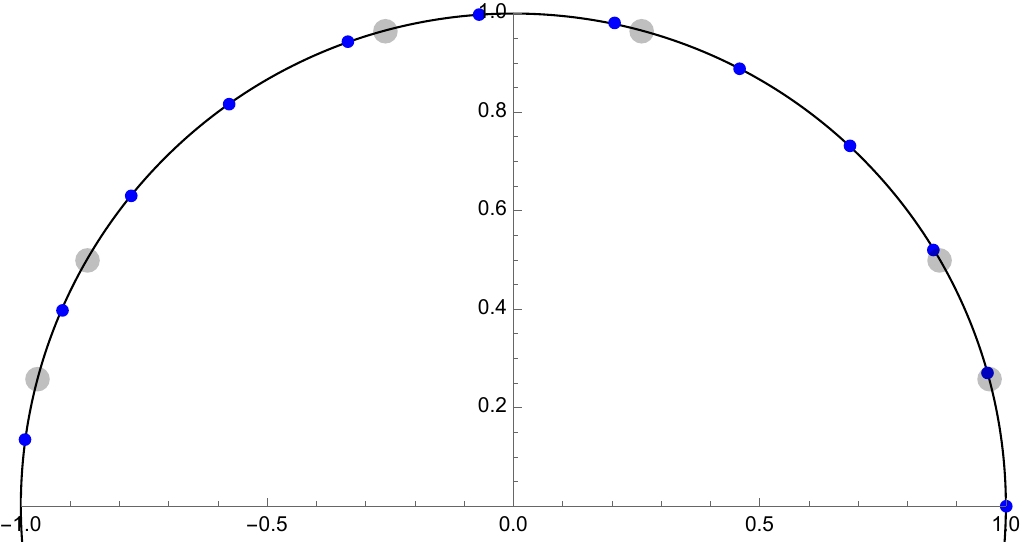}
    \caption{Plot of $x$-values: $x_{\alpha}$ (blue dots) and zeros of the Alexander polynomial $\{x\,\vert\, \Delta_{T(3,4)}(x^2)=0\}$ (gray dots) for $S^3_{+1/2}(T(3,4)) = \overline{\Sigma(3,4,23)}$}
    \label{fig:3_4_23}
\end{figure}

\paragraph{} Similar estimate should hold for $+1/r$-surgeries as well. Consider, for example, $+1/2$-surgery on $T(3,4)$ which gives reverse-oriented Brieskorn sphere $\overline{\Sigma(3,4,23)}$.
In Table \ref{table:sol_table_s34231}, we have listed the 12 zeros of the Alexander polynomial $\Delta_{T(3,4)}(x^2)$
as well as the intersection set $\{(\mathcal{A}_{T(3,4)}^{ \mathrm{irred}}(x,y)\equiv 1+x^{12}y=0) \cap (xy^2=1)\}$. These set of points are shown in Figure \ref{fig:3_4_23}. In this case one can deduce the degree $\deg \sigma = 3$, and we take this number as the estimate for the number of flat connections at infinity.

\begin{table}[h!]
    \centering
    \small
    \renewcommand{\arraystretch}{1.25} 
    \setlength{\tabcolsep}{0.1cm}        
    \begin{tabular}{||p{7.2cm}|p{7.2cm}||}
    \hline
       \textbf{$x$ : $\Delta_{T(3,4)}(x^2)=0$} 
       & \textbf{$\{x_\alpha\}$ : $\mathcal{A}_{T(3,4)}^{\mathrm{irred}}(x,y)=0 \,\cap\, xy^2=1$} 
       \\ \hline\hline     
       \begin{tabular}[c]{@{}l@{}}
       $-0.965926 \pm 0.258819 i$, 
       $-0.866025 \pm 0.500000 i$, \\
       $-0.258819 \pm 0.965926 i$, 
        $-0.258819 \pm 0.965926 i$, \\
        $-0.866025 \pm 0.500000 i$, 
        $-0.965926 \pm 0.258819 i$
       \end{tabular}
       &
       \begin{tabular}[c]{@{}l@{}}
       $0.682553 \pm 0.730836 i$, 
       $0.203456 \pm 0.979084 i$, \\
       $-0.990686 \pm 0.136167 i$, 
        $-0.962917 \pm 0.269797 i$, \\
       $-0.334880 \pm 0.942261 i$, 
       $-0.775711 \pm 0.631088 i$, \\
    $-0.460065 \pm 0.887885 i$, 
       $-0.068242 \pm 0.997669 i$, \\
       $-0.576680 \pm 0.816970 i$, 
        $-0.854419 \pm 0.519584 i$, \\
       $-0.917211 \pm 0.398401 i$
       \end{tabular}
       \\ \hline
    \end{tabular}
    \caption{ $x$: $\Delta_{T(3,4)}(x^2)=0$ and the $\{x_\alpha\}$  : $\mathcal{A}_{T(3,4)}^{\mathrm{irred}}(x,y)=0 \,\cap\, xy^2=1$}
    \label{table:sol_table_s34231}
\end{table}

Note that the path method gives all $33$ finite flat connections:
\begin{equation}\label{eq:flat connections 3,4,23}
\{CS_\alpha\} =
\left\{
\begin{aligned}
&\tfrac{11}{276},\ \tfrac{143}{1104},\ \tfrac{191}{1104},\ \tfrac{263}{1104},
 \tfrac{287}{1104},\ \tfrac{83}{276},\ \tfrac{359}{1104},\ \tfrac{383}{1104}, \\
&\tfrac{107}{276},\ \tfrac{431}{1104},\ \tfrac{479}{1104},\ \tfrac{503}{1104},
 \tfrac{527}{1104},\ \tfrac{551}{1104},\ \tfrac{143}{276},\ \tfrac{155}{276}, \\
&\tfrac{695}{1104},\ \tfrac{743}{1104},\ \tfrac{191}{276},\ \tfrac{203}{276},
 \tfrac{815}{1104},\ \tfrac{839}{1104},\ \tfrac{227}{276},\ \tfrac{911}{1104}, \\
&\tfrac{935}{1104},\ \tfrac{983}{1104},\ \tfrac{251}{276},\ \tfrac{1031}{1104},
 \tfrac{263}{276},\ \tfrac{1055}{1104},\ \tfrac{1079}{1104},\ \tfrac{275}{276},\ \tfrac{1103}{1104}
\end{aligned}
\right\} \mod 1.
\end{equation}

Now we proceed to the Chern-Simons action corresponding to flat connections at infinity. We consider the integral \eqref{CSaction}
 and the path from every solution $x_0\in \Delta_0^+$ to the point  $x_{\alpha^*}=1$ . This gives three values
 \begin{equation}
 \{CS_{\alpha^*}\}=\{\frac{1}{48},\frac{1}{40},\frac{1}{12}\} \mod 1.
 \end{equation}
  As a result, for the Brieskorn sphere $\overline{\Sigma(3,4,23)}$ we obtained 33 finite flat connections and 3 flat connections at infinity.\footnote{It would be very interesting to find any parallels with other studies of inverse-oriented Brieskorn spheres \cite{Harichurn:2025suf,Adams:2025qgj}.}

We have performed analogous calculation for $-\frac{1}{2}$-surgery on $T(3,4)$, i.e. $\Sigma(3,4,25)$. We obtained three flat connections at infinity with Chern-Simons actions
 \begin{equation}
 \{CS_{\alpha^*}\}=\{\frac{23}{48}, \frac{11}{12}, \frac{47}{48}\} \mod 1.
 \end{equation}



\paragraph{}
We will now investigate the presence of flat connections at infinity for non-torus knots. In particular, we will focus on  $\pm 1/r$ surgery on a twist and double twist knot.

\subsection{Twist knot family}

\begin{figure}[h!]
    \centering  \includegraphics[width=0.75\linewidth]{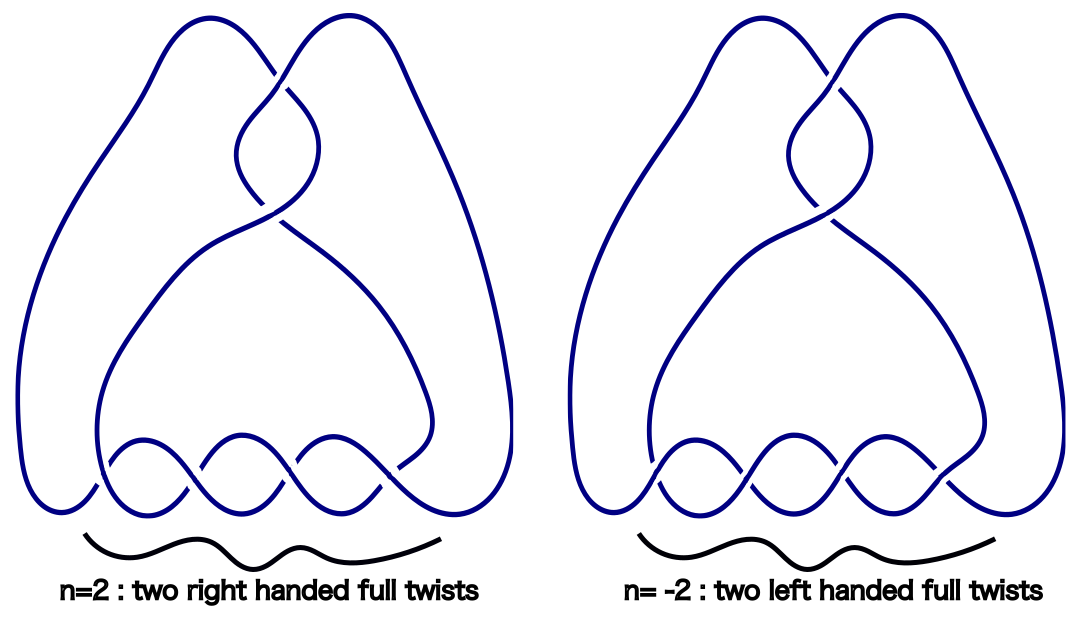}
    \caption{Twist knot $K_n$ with $n=2$ and $n=-2$}
\label{Twist knot}
\end{figure}

The twist knot family $K_n$ (where $n\in\mathbb{Z}\setminus \{0\}$ denotes the number of full twists) is naturally classified into two subfamilies (see Figure \ref{Twist knot}):

\begin{itemize}
  \item $n > 0$ implies \textit{right-handed} full twists, the corresponding knots are $~3_1,~5_2,~7_2,~9_2$ for $n=1,2,3,4$ respectively.
  \item $n < 0$ implies  \textit{left-handed} full twists, the corresponding knots are $4_1,~6_1,~8_1,~10_1$ for $n=-1,-2,-3,-4$ respectively.
\end{itemize}

\subsubsection*{\textbf{Case 1}: $n>0$ } 
The fundamental group $\pi_1(S^3 \setminus K_n)$ with  $n>0$  has two generators, $a$ and $b$, which satisfy the  relation:
\begin{equation}\label{fundamental twist}
(b a^{-1} b^{-1} a)^n a \; = \; b (b a^{-1} b^{-1} a)^n
\end{equation}
 In this case, $\mu = a$ represents the meridian, and the longitude is given by 
$$\lambda = (a b^{-1} a^{-1} b)^n (b a^{-1} b^{-1} a)^n.$$

\noindent Following the procedure from Section \ref{sec: counting}, we compute the  total number of flat connections at infinity under rational $\pm1/r$  surgery. In case of twist knot $K_n$ with $n > 0$, we find the number of flat connections at infinity to be
$$
\text{\# Flat connections at $\infty$}  =
\begin{cases}
2n - 1, & \text{if } r \in 2\mathbb{Z} \\[0.5em]
0, & \text{if } r \in 2\mathbb{Z} + 1
\end{cases}
$$
\noindent We present the corresponding $C$-polynomials, the number of flat connections at infinity, and the parameter $C$ for various twist knots in Table \ref{Tabletwist1}. In particular, this recovers the previous result from Section \ref{3_1 example} for the trefoil knot.

\begin{table}
\centering
\begin{adjustbox}{width=\textwidth}
\begin{tabular}{||c|>{\raggedright\arraybackslash}p{0.3\textwidth}|>{\raggedright\arraybackslash}p{0.45\textwidth}|c||}
\hline
Knot & $C$ Polynomial  & Zeros of $C$ polynomial & \# $\alpha^*$ \\ \hline\hline

$3_1$ & 
C+1 & 
$-1$ & 1 \\ \hline

$5_2$ & 
$C^3+ C^2+2 C+1$ & 
$-0.56984,\newline -0.21508 \pm 1.30714i$ & 3 \\ \hline

$7_2$ & 
$C^5+ C^4+4 C^3+3 C^2+3 C+1$ & 
$-0.416284,\newline -0.233677 \pm 0.885557i,\newline -0.0581814 \pm 1.69128i$ & 5 \\ \hline

$9_2$ & 
$C^7+ C^6+6 C^5+5 C^4+10 C^3+6 C^2+4 C+1$ & 
$-0.333643,\newline -0.216837 \pm 0.664672i,\newline -0.0931703 \pm 1.35015i,\newline -0.023171 \pm 1.82953i$ & 7 \\ \hline
\end{tabular}
\end{adjustbox}
\caption{Flat connections at infinity for twist knots ($n > 0$) under rational Dehn surgery.}
\label{Tabletwist1}
\end{table}

\subsubsection*{\textbf{Case 2}: $n<0$ }
For $n < 0$, the fundamental group of the twist knots, expressed in terms of the Wirtinger presentation, is given by
\begin{equation}
\label{fundamental twist neg}
(a b^{-1} a^{-1} b)^{|n|} a \; = \; b (a b^{-1} a^{-1} b)^{|n|}
\end{equation}
The meridian and longitude are $\mu = a$, and  
$\lambda = (b a^{-1} b^{-1} a)^{|n|} (a b^{-1} a^{-1} b)^{|n|}$ respectively.
%
%
For $n < 0$, Table~\ref{Tabletwist2} shows that
$$
\text{\# Flat connections at $\infty$} =
\begin{cases}
0, & \ r \in 2\mathbb{Z} + 1 \\[0.4em] 2|n|, & \ r \in 2\mathbb{Z}
\end{cases}
$$
In Table~\ref{Tabletwist2}, we compile the $C$-polynomials, the values of the parameter $C$, and the corresponding number of flat connections at infinity for several twist knots.

\begin{table}[H]
\centering
\begin{adjustbox}{width=\textwidth}
\begin{tabular}{||c|>{\raggedright\arraybackslash}p{0.3\textwidth}|>{\raggedright\arraybackslash}p{0.45\textwidth}|c||}
\hline
Knot & $C$ Polynomial  & Zeros of $C$ polynomial & \# $\alpha^*$ \\ \hline\hline

$4_1$ & 
$1 -  C + C^2$ & 
$0.5 \pm 0.866025i$ & 2 \\ \hline

$6_1$ & 
$C^4- C^3+3 C^2-2 C+1$ & 
$0.104877 \pm 1.55249i,\newline 0.395123 \pm 0.506844i$ & 4 \\ \hline

$8_1$ & 
$C^6-1 C^5+5C^4-4 C^3+6 C^2-3 C+1$ & 
$0.035468 \pm 1.7753i,\newline 0.142924 \pm 1.15952i,\newline 0.321608 \pm 0.359079i$ & 6 \\ \hline

$10_1$ & 
$C^8- C^7+7 C^6-6 C^5+15 C^4-10 C^3+10 C^2-4 C+1$ & 
$0.0159535 \pm 1.86641i,\newline 0.0640347 \pm 1.48479i,\newline 0.147789 \pm 0.913548i,\newline 0.272222 \pm 0.278653i$ & 8 \\ \hline
\end{tabular}
\end{adjustbox}
\caption{Flat connections at infinity for twist knots ($n < 0$) under rational Dehn surgery.}
\label{Tabletwist2}
\end{table}

Let us illustrate the procedure of computing the Chern-Simons action for the two flat connections at infinity 
for surgery manifold $S^3_{-1/2}(4_1)$.
The irreducible $A$-polynomial is  

\begin{equation}
\mathcal{A}^{\text{irred}}_{4_1}(x,y) 
= -x^{4} + (1 - x^{2} - 2x^{4} - x^{6} + x^{8})y - x^{4}y^{2}.
\end{equation}
Note that, the $\mathcal{A}$-polynomial is quadratic in $y$. Hence, we need to use a two sheeted branch cover   $y_{1}(x)$ and $y_{2}(x)$.  
The Alexander polynomial of the $K_{-1}=4_1$ knot is  
\begin{equation}
\Delta_{4_1}(x^{2}) = -x^{2} - x^{-2} + 3\,.
\end{equation}
The roots of $\Delta_{4_1}(x^{2})$ are $\{x_a\}=\left\{\frac{1}{2} \left(-1-\sqrt{5}\right),\frac{1}{2} \left(1-\sqrt{5}\right),\frac{1}{2} \left(\sqrt{5}-1\right),\frac{1}{2} \left(1+\sqrt{5}\right)\right\}$.
In order to compute the Chern-Simons action corresponding to the singular point $(x,y)=(1,-1)$, we can take a straight-line path from the root $x_0=\frac{1}{2} \left(\sqrt{5}-1\right)$ of $\Delta_{4_1}(x^{2})$ 
to $x_{\alpha^*}=1$. Taken along the first branch $y_1(x)$, it gives $CS_{\alpha^*}[S_{-1/2}^3(4_1)] =0.5 + 0.0514175 ~i$.  
\begin{figure}
    \centering
\includegraphics[width=0.8\linewidth]{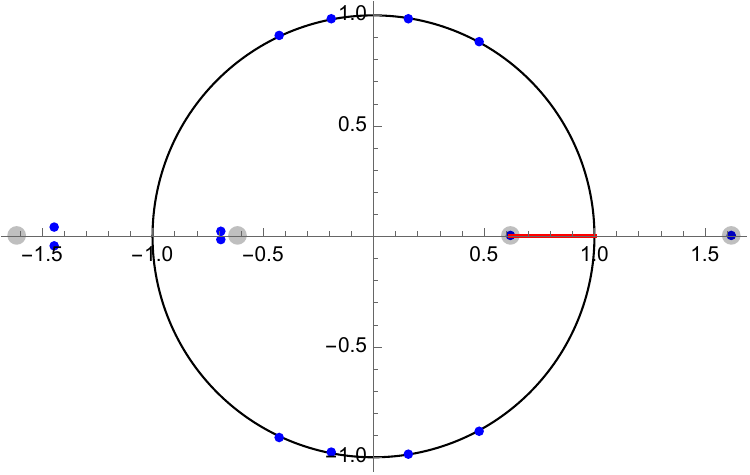}
    \caption{Plot of $x$-values: $x_{\alpha}$ (blue dots), $\{x_a\vert \Delta_{4_1}(x^2)=0\}$ (gray dots) and the path (red line) connecting $x_{\alpha^*}=1$ and $x_a=x_0$ for surgery manifold $S^3_{-1/2}(4_1)$}
    \label{fig:placeholder}
\end{figure}
The Chern-Simons action computed along the other branch $y_2(x)$ is $0.5 - 0.0514175 ~i$.

For the twist knot family, since these surgery manifolds for generic $r$ are not plumbed 3-manifolds, we can only rely on the path-method. As a result, we have verified that the number of $C$-values matches  the number of flat connections at infinity obtained using path method for $|n|< 5$.
Note also that twist knots $K_n$ can be viewed as a part of a larger family of double twist knots denoted $J(s=2,2n)$, where both $s$ and $2n$ denote the total number of half-twists. 
\begin{figure}[t!]
    \centering  \includegraphics[width=0.75\linewidth]{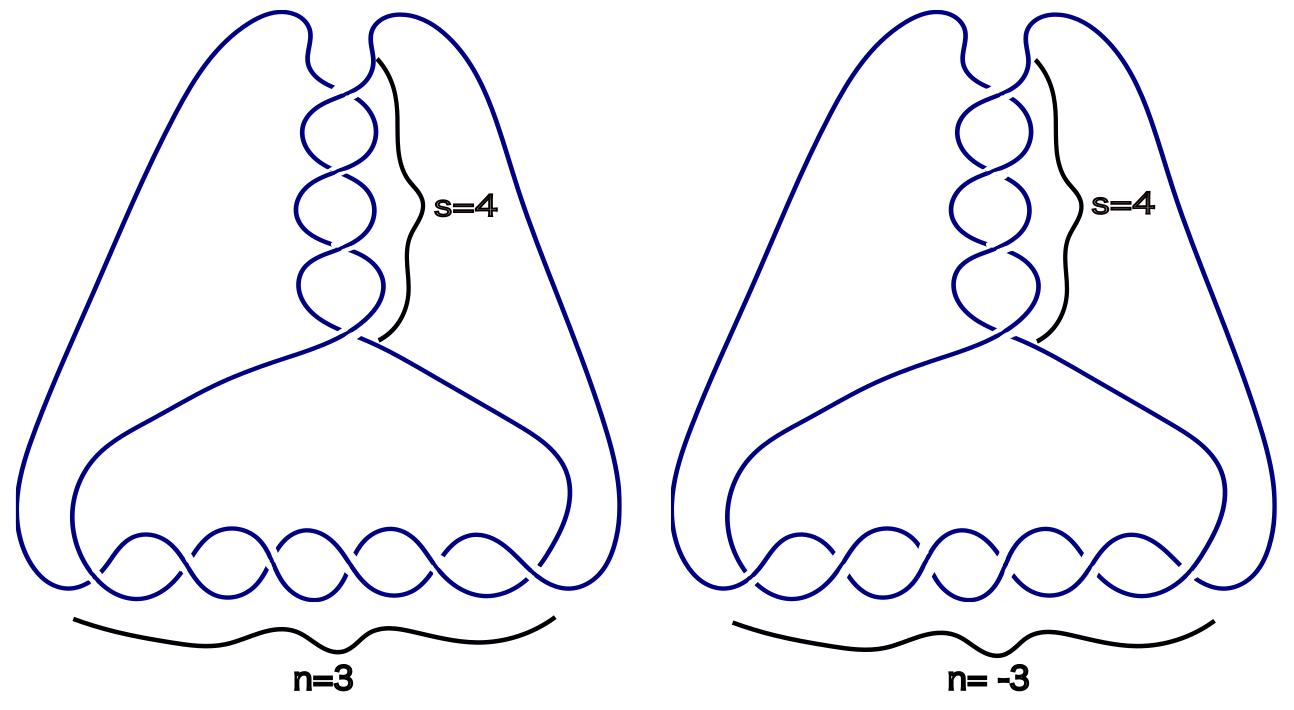}
    \caption{Double twist knots: $J(s=4,2n=6)$, $J(s=4,2n=-6)$. Here, $s=4$ denotes four right handed vertical half twists; $2n=6,-6$ specifies six right handed, left handed horizontal half twists respectively}
\label{Double Twist knot}
\end{figure}
 Next we will investigate flat connections at infinity and the corresponding Chern-Simons action for more general knots of type $J(s,2n)$ (see Figure \ref{Double Twist knot}) and their rational surgeries.

\subsection{Double twist knot family}

The fundamental group of the complement of the  double twist knot $J(s, 2n)$ in $S^3$ is  \cite{hoste2004formula}:
\begin{equation}
  \pi_1(S^3\setminus J(s, 2n)) = \langle a, b \mid a {w_s^n} = w_s^{n} b \rangle\,,  
\end{equation}
where $w_s$ is defined as:
$$
w_s = 
\begin{cases}
(ab^{-1})^m (a^{-1}b)^m & \text{if } s = 2m, \\
(ab^{-1})^m ab (a^{-1}b)^m & \text{if } s = 2m + 1.
\end{cases}
$$
 The meridian $\mu$ of the knot corresponds to the generator $a$, while the longitude $\lambda$ is given by:
\begin{equation}
  \lambda = w_s^n (w_s^n)^{-1} a^{-2\varepsilon(s,n)},  
\end{equation}
%
where $\varepsilon(s, n) = 0$ for even $s$ and $\varepsilon(s, n) = 2n$ for odd $s$.
We compute the $A$-polynomials for this family of knots using the procedure provided in \cite{petersen2015polynomials}.
We then follow the same algorithm outlined in Section \ref{sec: counting} as the first approach, in order to compute the number of flat connections at infinity.
For this family of knots, we found that this number is expressed in terms of $m$ and $n$ as follows:
\begin{equation}
\text{\# Flat connections at $\infty$} =
\begin{cases}
\lvert 2mn \rvert, & \text{if } s = 2m,\ n < 0 \\[0.5em]
sm - 1, & \text{if } s = 2m + 1,\ n > 0 \\[0.5em]
\lvert sn \rvert, & \text{if } s = 2m + 1,\ n < 0
\end{cases}
\end{equation}
We also tabulate the $C$-polynomial, its roots and the corresponding number of flat connections at infinity for $J(s, 2n)$ in Tables \ref{table:DoubleTwistKnots_J3_2n_pos}, \ref{table:DoubleTwistKnots_J3_2n_neg}, \ref{table:DoubleTwistKnots_J4_2n_pos} and \ref{table:DoubleTwistKnots_J4_2n_neg}.

\paragraph{}
\begingroup
\footnotesize

\begin{longtable}{||c|p{0.65\linewidth}|c||}
\hline
\textbf{Knot} & \textbf{$C$-Polynomial \& its zeros} & \textbf{\# $\alpha^*$} \\
\hline\hline
\endfirsthead

\multicolumn{3}{c}{\bfseries Table \thetable\ continued}\\
\hline
\textbf{Knot} & \textbf{$C$-Polynomial \& its zeros} & \textbf{\# $\alpha^*$} \\
\hline
\endhead

\hline \multicolumn{3}{r}{Continued on next page}\\ \hline
\endfoot

\hline
\endlastfoot

$J(3,2)$ &
$1 - C + C^2$ \newline
Zeros: $0.5000\pm0.8660\,i$ & 2 \\ \hline

$J(3,4)$ &
$1 - C - C^2 + 4C^3 - 3C^4 + C^5$ \newline
Zeros: $-0.5880,\;0.5613\pm0.5578\,i,\;1.2327\pm1.0938\,i$ & 5 \\ \hline

$J(3,6)$ &
$1 - 6C^2 + 10C^3 - C^4 - 10C^5 + 11C^6 - 5C^7 + C^8$ \newline
Zeros: $-0.8366,\;-0.3348,\;0.6073\pm0.4144\,i,\;1.0821\pm0.8777\,i,\;1.3964\pm1.2056\,i$ & 8 \\ \hline

$J(3,8)$ &
$1 + 2C - 13C^2 + 11C^3 + 23C^4 - 48C^5 + 24C^6 + 20C^7 - 35C^8 + 22C^9 - 7C^{10} + C^{11}$ \newline
Zeros: $-0.9123,\;-0.6491,\;-0.2061,\;0.6449\pm0.3264\,i,\;1.0073\pm0.7329\,i,\;1.2849\pm1.0517\,i,\;1.4466\pm1.2541\,i$ & 11 \\ \hline

$J(3,10)$ &
$1 + 5C - 20C^2 - 4C^3 + 86C^4 - 98C^5 - 46C^6 + 188C^7 - 147C^8 - 7C^9 + 98C^{10} - 84C^{11} + 37C^{12} - 9C^{13} + C^{14}$ \newline
Zeros: $-0.9454,\;-0.7818,\;-0.5097,\;-0.1353,\;-0.05398,\;0.6769\pm0.2667\,i,\;0.9653\pm0.6300\,i,\;1.2056\pm0.9254\,i,\;1.3706\pm1.1445\,i,\;1.4678\pm1.2781\,i$ & 14 \\ \hline

$J(3,12)$ &
$1 + 9C - 24C^2 - 48C^3 + 190C^4 - 74C^5 - 390C^6 + 592C^7 - 77C^8 - 605C^9 + 678C^{10} - 204C^{11} - 213C^{12} + 285C^{13} - 165C^{14} + 56C^{15} - 11C^{16} + C^{17}$ \newline
Zeros: $-0.9627,\;-0.8511,\;-0.6659,\;-0.4079,\;-0.0937,\;0.7043\pm0.2240\,i,\;0.9397\pm0.5528\,i,\;1.1492\pm0.8251\,i,\;1.3052\pm1.0416\,i,\;1.4139\pm1.1974\,i,\;1.4785\pm1.2914\,i$ & 17 \\
\hline

\caption{For double twist Knot $J(3,2n)$ with $n>0:$ $C$-polynomial \& its zeros, and number of flat connections at infinity}
\label{table:DoubleTwistKnots_J3_2n_pos}
\end{longtable}
\endgroup

\begingroup
\footnotesize

\begin{longtable}{||c|p{0.65\linewidth}|c||}
\hline
\textbf{Knot} & \textbf{$C$-Polynomial \& its zeros} & \textbf{\# $\alpha^*$} \\
\hline\hline
\endfirsthead

\multicolumn{3}{c}{\bfseries Table \thetable\ continued}\\
\hline
\textbf{Knot} & \textbf{$C$-Polynomial \& its zeros} & \textbf{\# $\alpha^*$} \\
\hline
\endhead

\hline \multicolumn{3}{r}{Continued on next page}\\ 
\endfoot

\hline
\endlastfoot

$J(3,-2)$ &
$1 + 2C - 3C^2 + C^3$ \newline
Zeros: $-0.324718,\;1.66236\pm0.56228\,i$ & 3 \\ \hline

$J(3,-4)$ &
$1 + 5C - 6C^2 - 4C^3 + 9C^4 - 5C^5 + C^6$ \newline
Zeros: $-0.762503,\;-0.170693,\;1.46202\pm1.04357\,i,\;1.50458\pm0.342767\,i$ & 6 \\ \hline

$J(3,-6)$ &
$1 + 9C - 6C^2 - 24C^3 + 31C^4 + C^5 - 25C^6 + 20C^7 - 7C^8 + C^9$ \newline
Zeros: $-0.884324,\;-0.55322,\;-0.107184,\;1.39388\pm0.804887\,i,\;1.413\pm0.251868\,i,\;1.46548\pm1.19399\,i$ & 9 \\ \hline

$J(3,-8)$ &
$1 + 14C + C^2 - 67C^3 + 50C^4 + 80C^5 - 136C^6 + 44C^7 + 57C^8 - 70C^9 + 35C^{10} - 9C^{11} + C^{12}$ \newline
Zeros: $-0.932035,\;-0.732484,\;-0.41908,\;-0.0738364,\;1.33959\pm0.652562\,i,\;1.35342\pm0.201153\,i,\;1.40995\pm1.0281\,i,\;1.47575\pm1.24977\,i$ & 12 \\ \hline

$J(3,-10)$ &
$1 + 20C + 20C^2 - 136C^3 + 14C^4 + 344C^5 - 334C^6 - 178C^7 + 517C^8 - 308C^9 - 70C^{10} + 214C^{11} - 147C^{12} + 54C^{13} - 11C^{14} + C^{15}$ \newline
Zeros: $-0.955372,\;-0.82314,\;-0.609051,\;-0.329052,\;-0.0539813,\;\newline 1.29867\pm0.549094\,i,\;1.31112\pm0.168411\,i,\;1.36006\pm0.893453\,i,\;1.43281\pm1.13424\,i,\;1.48265\pm1.27597\,i$ & 15 \\ \hline

$J(3,-12)$ &
$1 + 27C + 57C^2 - 224C^3 - 178C^4 + 914C^5 - 352C^6 - 1408C^7 + 1811C^8 - 7C^9 - 1695C^{10} + 1500C^{11} - 250C^{12} - 537C^{13} + 540C^{14} - 264C^{15} + 77C^{16} - 13C^{17} + C^{18}$ \newline
Zeros:$-0.968479,\;-0.874676,\;-0.721067,\;-0.513053,\;-0.265491,$\;\newline$-0.0411716,\;1.26726\pm0.474469\,i,\;1.27925\pm0.145358\,i,\;1.319\pm0.787833\,i,\;1.39004\pm1.02614\,i,\;1.44932\pm1.192\,i,\;1.48711\pm1.29027\,i$ & 18 \\
\hline

\caption{For double Twist Knots $J(3,2n)$ with $n<0:$ $C$-polynomial and its zeros, and number of flat connections at infinity }
\label{table:DoubleTwistKnots_J3_2n_neg}
\end{longtable}
\endgroup

\begingroup
\footnotesize
\begin{longtable}{||c|p{0.65\linewidth}|c||}
\hline
\textbf{Knot} & \textbf{$C$-Polynomial \& its zeros} & \textbf{\# $\alpha^*$} \\
\hline\hline
\endfirsthead

\multicolumn{3}{c}{\bfseries Table \thetable\ continued}\\
\hline
\textbf{Knot} & \textbf{$C$-Polynomial \& its zeros} & \textbf{\# $\alpha^*$} \\
\hline
\endhead

\hline \multicolumn{3}{r}{Continued on next page}\\ \hline
\endfoot

\hline
\endlastfoot

$J(4,2)$ &
$1 + 2C - 3C^2 + C^3$ \newline
Zeros: $-0.324718,\;1.662359\pm0.562280\,i$ & 3 \\ \hline

$J(4,4)$ &
$1 + 4C - 2C^2 + 6C^3 - 19C^4 + 18C^5 - 7C^6 + C^7$ \newline
Zeros: $-0.205569,\;2.102785\pm0.665457\,i,\;-0.192440\pm0.547877\,i,\;1.692440\pm0.318148\,i$ & 7 \\ \hline

$J(4,6)$ &
$1 + 6C + 3C^2 + 23C^3 - 61C^4 + 72C^5 - 116C^6 + 156C^7 - 119C^8 + 50C^9 - 11C^{10} + C^{11}$ \newline
Zeros: $-0.156607,\;-0.228082\pm0.691240\,i,\;-0.139347\pm0.388084\,i,\;1.739636\pm0.215754\,i,\;2.002351\pm0.529597\,i,\;2.203745\pm0.721451\,i$ & 11 \\ \hline

$J(4,8)$ &
$1 + 8C + 12C^2 + 60C^3 - 114C^4 + 212C^5 - 558C^6 + 866C^7 - 1051C^8 + 1260C^9 - 1238C^{10} + 834C^{11} - 363C^{12} + 98C^{13} - 15C^{14} + C^{15}$ \newline
Zeros: $-0.128631,\;-0.246183\pm0.735527\,i,\;-0.176952\pm0.577289\,i,\;-0.111139\pm0.298099\,i,\;1.776498\pm0.161702\,i,\;1.956618\pm0.432309\,i,\;2.129278\pm0.629732\,i,\;2.236197\pm0.747339\,i$ & 15 \\ \hline

$J(4,10)$ &
$1 + 10C + 25C^2 + 125C^3 - 150C^4 + 552C^5 - 1684C^6 + 2940C^7 - 5177C^8 + 8378C^9 - 10691C^{10} + 11777C^{11} - 11769C^{12} + 9744C^{13} - 6056C^{14} + 2680C^{15} - 815C^{16} + 162C^{17} - 19C^{18} + C^{19}$ \newline
Zeros: $-0.110151,\;-0.255197\pm0.754725\,i,\;-0.207363\pm0.658054\,i,\;-0.140294\pm0.488427\,i,\;-0.093824\pm0.241653\,i,\;1.804251\pm0.128899\,i,\;1.934928\pm0.362335\,i,\;2.077555\pm0.550429\,i,\;2.184871\pm0.682757\,i,\;2.250149\pm0.760513\,i$ & 19 \\
\hline

\caption{For double twist Knot $J(4,2n)$ with ($n>0$): $C$-polynomial \& its zeros, and number of flat connections at infinity}
\label{table:DoubleTwistKnots_J4_2n_pos}
\end{longtable}
\endgroup

\begingroup
\footnotesize
\begin{longtable}{||c|p{0.65\linewidth}|c||}
\hline
\textbf{Knot} & \textbf{$C$-Polynomial \& its zeros} & \textbf{\# $\alpha^*$} \\
\hline\hline
\endfirsthead

\multicolumn{3}{c}{\bfseries Table \thetable\ continued}\\
\hline
\textbf{Knot} & \textbf{$C$-Polynomial \&  its zeros } & \textbf{\# $\alpha^*$} \\
\hline
\endhead

\hline \multicolumn{3}{r}{Continued on next page}\\ \hline
\endfoot

\hline
\endlastfoot

$J(4,-2)$ &
$1 - 2C + 7C^2 - 5C^3 + C^4$ \newline
Zeros: $0.100768\pm0.400532\,i,\;2.399232\pm0.325640\,i$ & 4 \\ \hline

$J(4,-4)$ &
$1 - 4C + 18C^2 - 22C^3 + 39C^4 - 50C^5 + 31C^6 - 9C^7 + C^8$ \newline
Zeros: $-0.165973\pm0.650629\,i,\;0.116795\pm0.251218\,i,\;2.251259\pm0.604856\,i,\;2.297919\pm0.193671\,i$ & 8 \\ \hline

$J(4,-6)$ &
$1 - 6C + 33C^2 - 59C^3 + 166C^4 - 264C^5 + 324C^6 - 396C^7 + 365C^8 - 210C^9 + 71C^{10} - 13C^{11} + C^{12}$ \newline
Zeros: $-0.224163\pm0.720432\,i,\;-0.086060\pm0.517646\,i,\;0.108247\pm0.181533\,i,\;2.212638\pm0.456283\,i,\;2.240962\pm0.140553\,i,\;2.248376\pm0.703111\,i$ & 12 \\ \hline

$J(4,-8)$ &
$1 - 8C + 52C^2 - 124C^3 + 450C^4 - 852C^5 + 1550C^6 - 2562C^7 + 3247C^8 - 3564C^9 + 3510C^{10} - 2722C^{11} + 1491C^{12} - 546C^{13} + 127C^{14} - 17C^{15} + C^{16}$ \newline
Zeros: $-0.244957\pm0.747590\,i,\;-0.165732\pm0.629537\,i,\;-0.042241\pm0.424718\,i,\;0.097796\pm0.141704\,i,\;2.182962\pm0.364200\,i,\;2.204514\pm0.111304\,i,\;2.212726\pm0.597582\,i,\;2.254933\pm0.739658\,i$ & 16 \\ \hline

$J(4,-10)$ &
$1 - 10C + 75C^2 - 225C^3 + 975C^4 - 2152C^5 + 5124C^6 - 10060C^7 + 16187C^8 - 24218C^9 + 31735C^{10} - 35237C^{11} + 34524C^{12} - 29584C^{13} + 20616C^{14} - 10904C^{15} + 4185C^{16} - 1122C^{17} + 199C^{18} - 21C^{19} + C^{20}$ \newline
Zeros: $-0.254660\pm0.760840\,i,\;-0.203455\pm0.683819\,i,\;-0.121619\pm0.551495\,i,\;-0.017131\pm0.358212\,i,\;0.088527\pm0.116089\,i,\;2.161005\pm0.302970\,i,\;2.178914\pm0.092584\,i,\;2.182856\pm0.512717\,i,\;2.225857\pm0.666508\,i,\;2.259706\pm0.756602\,i$ & 20 \\
\hline

\caption{ For double Twist Knots $J(4,2n)$ with $(n<0)$: $C$-polynomial \& its zeros, and number of flat connections at infinity}
\label{table:DoubleTwistKnots_J4_2n_neg}
\end{longtable}
\endgroup

\section{Conclusions and future directions}\label{sec:conclusions}

In this work we have explored the infinite asymptotic ends of moduli spaces $\mathcal{M}_{\rm flat}^{\rm irr}$ of irreducible $SL(2,\mathbb{C})$ flat connections on 3-manifolds given by knot complements in $S^3$ and $\pm1/r$-surgeries on them, continuing the line of research in \cite{gukov2024categorification}. Such asymptotic flat connections are also called flat connections at infinity, and are one of the most mysterious and appealing features of complex Chern-Simons theory. We developed and successfully applied the two approaches to their study, outlined in Section~\ref{sec: counting}. First, we started with the Wirtinger presentation of the knot group and followed the algorithm to obtain the explicit form (in terms of $\pi_1$ representations) and the number of such flat connections by solving a system of algebraic equations. Second, we computed the Chern-Simons action corresponding to each of the flat connection, including those at infinity, and produced their count. As the main outcome, we confirmed the existence of flat connections at infinity for surgery manifolds given by $\pm 1/r$-surgery on torus and double twist knots. 
Supplementing these results, we also used (when it was applicable) the plumbed graph approach to verify the appearance of such asymptotic connections. Interestingly, our proposal with $P_a$ \eqref{ContFracMethod} computation  modified (that is, removing the  bivalent vertex with highest magnitude of the framing in the plumbing graph) indeed shows the presence of such flat connections. 
We also observed that the number of flat connections at infinity follows a pattern for various knot families, as shown in Section \ref{sec: case studies}.

The next natural step would be to observe the conditions when such flat connections at infinity contribute to the Chern-Simons partition function \eqref{eq:CS path integral}. For example, one of the most promising directions is to investigate the Borel plane for the reverse-oriented Brieskorn spheres $\overline{\Sigma(p,q,r)}$ in order to possibly detect such flat connections 
The above said also naturally fits into the theory of $\widehat{Z}$ invariants and 3d-3d correspondence \cite{Gukov:2017kmk}. Therefore, it would be very interesting to explore the implications of flat connections at infinity from this perspective, and to explain them from the point of view of 3d $\mathcal{N}=2$ theory $T[M]$ associated to a 3-manifold $M$.

\appendix

\section{Explicit calculation of the gauge connection}\label{gauge connection}
 We follow the construction proposed in \cite{kirk1990chern,kirk1993chern}, where developing maps are used to study the monodromy of flat $ \mathrm{SL}(2, \mathbb{C})$ connections and their relation to geometric structures on 3-manifolds.
If $ \rho: \pi_1(T^2) \to \mathrm{SL}(2, \mathbb{C}) $ be a representation of the fundamental group of the boundary of three-manifold, we work with the representation  that has properties where the trace of every element is 
$\mathrm{Tr}(\rho(\gamma)) =\pm 2 \quad \text{for all } \gamma \in \pi_1(T^2).$
This condition implies that all elements in the image of  $\rho$ are unipotent. In particular, the images of the standard generators, the meridian $\mu$ and the longitude $\lambda $ can be expressed in the form:
$$
\rho(\mu) = (-1)^u \begin{pmatrix} 1 & a_1 \\ 0 & 1 \end{pmatrix}, \qquad
\rho(\lambda) = (-1)^v \begin{pmatrix} 1 & b_1 \\ 0 & 1 \end{pmatrix},
$$
for some $a_1, b_1 \in \mathbb{C} $ and integers $ u, v \in \mathbb{Z} $.

To describe the associated holonomy, we consider a developing map
$
D : \mathbb{R}^2 \longrightarrow \mathrm{SL}(2, \mathbb{C}),
$
as
$$D({x}_1, {x}_2) =
\begin{pmatrix}
\exp\left( \frac{i}{2}(u {x}_1 + v{x}_2) \right) & -\frac{(a_1{x}_1 + b_1 x_2)}{2\pi} \exp\left( \frac{i}{2}(u{x}_1 + v{x}_2)\right) \\
0 & \exp\left( -\frac{i}{2}(u{x}_1 + v{x}_2) \right).
\end{pmatrix}
$$
This map satisfies an additional property namely,
$
D({x}_1 + 2\pi m, {x}_2 + 2\pi n) = D({x}_1, {x}_2) \cdot \rho(m\mu + n\lambda)^{-1}
\quad \text{for all } m, n \in \mathbb{Z}.
$
 In terms of the developing map, the flat connection one-form is defined as $A = -dD \cdot D^{-1}$ whose holonomy is given by the representation $\rho $.
An explicit computation yields
$$ A = \frac{1}{2\pi}
\begin{pmatrix}
\frac{-i u}{2} \, d{x}_1 - \frac{i v}{2} \, d{x}_2 &~~ \left( \frac{a_1}{2\pi} \, d{x}_1 + \frac{b_1}{2\pi} \, d{x}_2 \right) \exp(u{x}_1 + v{x}_2) \\
0 & \exp(\frac{-i u}{2} \, d{x}_1 - \frac{i v}{2} \, d{x}_2)
\end{pmatrix}.
$$
Furthermore, we can choose a gauge transformation matrix $g$ of the form :
$$g =
\begin{pmatrix}
\omega & 0 \\
0 & 1/\omega
\end{pmatrix}$$
to understand the behaviour of gauge connection $A$.

\section{Chern-Simons invariants from plumbing matrix}\label{plumbed methd}  
The three-manifolds constructed from surgery of a $r$-component framed link $[L, \mathbf f]$  where $\mathbf f=f_1,f_2,\ldots f_r$ is a $r$-tuple giving the framing numbers (or, self crossing number) of the component knots $K_1,K_2, \ldots K_r$ constituting the link $[L,\mathbf f]$. If the linking number between component knots satisfy $\ell k( K_i,K_j)= 1 ~{\rm or}~0~,$ then those framed links can be depicted as  a plumbing graph $\Gamma$ with $r$ vertices and edges connecting the vertices \cite{Gukov:2017kmk}. 

\begin{figure}[h!]
\centering
\includegraphics[width=0.7\linewidth]{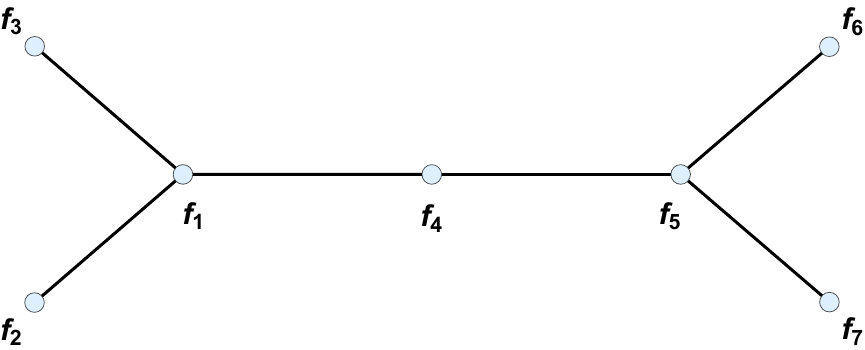}
\caption{Example of plumbed three-manifold.}
\label{fig:plumbed graph1}
\end{figure}

The information of $\Gamma$ are encoded in the adjacency matrix $\mathbf B^{(\Gamma)}$ and the corresponding manifolds $M_{\Gamma}$ are called {\it plumbed three-manifolds} (see Figure \ref{fig:plumbed graph1}). The $\mathbf B^{(\Gamma)}$ diagonal entries will be  $\mathbf{B}_{ii}^{(\Gamma)} = f_i $  and  the off-diagonal ($i \neq j $) entries will be
$$
\mathbf{B}_{ij}^{\Gamma} = 
\begin{cases}
1 & \text{if vertices } i \text{ and } j \text{ are connected by an edge}, \\
0 & \text{otherwise}.
\end{cases}
$$
Furthermore, the structure of the matrix $\mathbf{B}^{(\Gamma)}$ can be refined by arranging the vertices according to their valency: high-valency vertices ($\mathbf{h}$, connected to more than two vertices), low-valency vertices ($\boldsymbol{l}$, connected to a single vertex), and bi-valent vertices. We compute the inverse of this matrix and denote the blocks associated to the high-high, high-low and low-low vertex sectors of ${\mathbf{B}^{(\mathbf{\Gamma})}}^{-1}$ as  \cite{gukov2024categorification} 
\begin{equation}
\quad
\mathbf{C} := (\mathbf{B^{(\Gamma)}}^{-1})_{\mathbf{hh}}, \quad
\mathbf{D} := (\mathbf{B^{(\Gamma)}}^{-1})_{\mathbf{h}\boldsymbol{\ell}}, \quad
\mathbf{A} := (\mathbf{B^{(\Gamma)}}^{-1})_{\boldsymbol{ll}}.
\label{CDA}
\end{equation}
Note that, in defining the above blocks the subspace of ${\mathbf{B}^{(\mathbf{\Gamma})}}^{-1}$ that is generated by the bi-valent vertices remains untouched. The partition function of a plumbed 3-manifold whose $|\det~\mathbf{B^{(\Gamma)}}|=1$ can be written as an integral over the variables $\mathbf{v}$ associated with the high-valency vertices of the plumbing graph. Analytic continuation of the Chern-Simons partition function $Z(M) \xrightarrow{k ~\text{complex}} Z[M^\Gamma; k] $ is given by
\begin{equation}
Z[M^\Gamma; k] \propto \int d^{|h|}\mathbf{v} \; R(\mathbf{v}) \, \exp\left( 2\pi i k \, S(\mathbf{v}) \right).
\label{part fun}
\end{equation}

\begin{figure}[t!]
    \centering
    \includegraphics[width=0.8\linewidth]{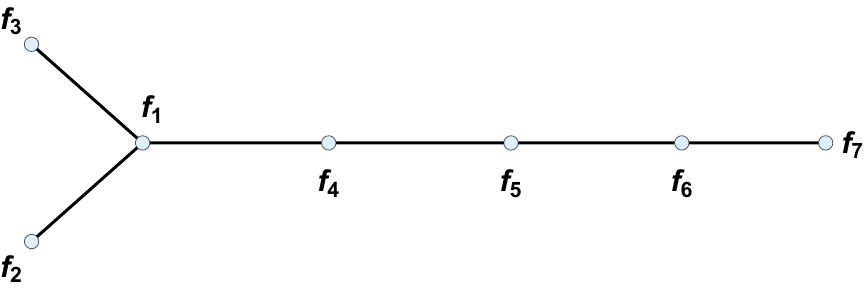}
    \caption{Plumbing graph with single high valency vertex}
    \label{fig:plumbing graph2}
\end{figure}

We can write the  explicit form of the integrands $S(\mathbf{v})$ and $R(\mathbf{v})$ as:
\begin{align}
 S(\mathbf{v}) = \frac{1}{4} \, \mathbf{v}^T \mathbf{C}^{-1} \mathbf{v}
\label{eq:S_function}   
\end{align}
\begin{equation}
 R(\mathbf{v}) = \frac{\prod_{a \in \boldsymbol{\ell}} \sin\left( \pi \, (\mathbf{v}^T \mathbf{C}^{-1} \mathbf{D} )_a \right)}
{\prod_{I \in \mathbf{h}} \left( \sin\left( \pi v_I \right) \right)^{\deg(I) - 2}}=\frac{ \prod_{a\in \boldsymbol{\ell} }\sin \left( \frac{\pi v_{h(a)} }{P_{a}} \right) }{\prod_{I \in \mathbf{h}} \left( \sin\left( \pi v_I \right) \right)^{\deg(I) - 2}},
\label{eq:R_function}   
\end{equation}
where $h(a)\in \mathbf{h}$ is a high-valency vertex that is most adjcent to the low-valency vertex $a$. In other words, the intermediate vertices, if any, connecting $a$ and $h(a)$ are all of valency 2. The integers $P_a$ are determined as the numerator of the following continued fraction 
\begin{align}
    f_a-\frac{1}{f_{i_1}- \frac{\textstyle 1}{\textstyle f_{i_2} -\frac{ \textstyle\cdots}{ \textstyle \cdots -\frac{ \textstyle 1}{\textstyle f_{i_s}}} } }&=-\frac{P_a}{Q_a},\label{ContFracMethod}
\end{align}
where $(f_a,~f_{i_1},~f_{i_2},\cdots, f_{i_s},~f_{h(a)})$ are the framings associated to the sequence of adjacent vertices $(a,~i_1,~i_2,\cdots, i_s,~h(a))$ and the integer $Q_a$ is coprime to $P_a$.  For illustration purpose, let us consider the case of plumbed three-manifolds whose plumbing graph contains a single high-valent vertex as shown in Figure \ref{fig:plumbing graph2}.
Such manifolds are alternatively  viewed as Seifert fibred three-manifolds with three singular fibres \cite{Cheng:2022rqr}. For this particular case, the function $R(\mathbf{v})\equiv R(v_1)$ takes the following form 
\begin{equation}
R({v}_1) = \frac{\prod_{a \in \boldsymbol{\ell} } \sin\left( \frac{\pi \, v_1 }{P_a}\right)}
{  \sin\left( \pi v_1 \right ) }.
\label{eq:R1_function}
\end{equation}
The product $a$ runs over the low-valency vertices with framings $f_2$, $f_3$ and $f_7$ in the aforementioned figure. The value of $P_a$ corresponding to each of these low valency vertices can be computed using the method described in (\ref{ContFracMethod}). As an example, for the low valent vertex $v_7$ with framing $f_7$ in Figure \ref{fig:plumbing graph2}, $P_7$ is obtained in the following way:
\begin{equation}
    f_7-\frac{1}{f_6-\frac{1}{f_5-\frac{1}{f_4}}}=-\frac{P_7}{Q_7},
    \label{P_a values}
\end{equation}
where the integers $P_7$ and $Q_7$ are assumed to be coprime. 

If $\mathbf{h}$ is the set of all high-valency vertices, we can choose a subset $\mathbf{h_1} \subset \mathbf{h}$, and denote its complement by $\mathbf{h_2} = \mathbf{h} / \mathbf{h_1}$. In this setting, the matrix $\mathbf{C}^{-1}$ admits the following block decomposition:
\begin{equation}
\mathbf{C}^{-1} =
\begin{bmatrix}
\tilde{A} & \tilde{D}^T \\
\tilde{D} & \tilde{C}
\end{bmatrix},
\qquad
\tilde{A} = (\mathbf{C}^{-1})_{\mathbf{h_1} \mathbf{h_1}}, \quad
\tilde{D}^{T} = (\mathbf{C}^{-1})_{\mathbf{h_1} \mathbf{h_2}}, \quad
\tilde{C} = (\mathbf{C}^{-1})_{\mathbf{h_2} \mathbf{h_2}}
\label{eq:Cinv_block_named}
\end{equation}
By analyzing the integral \eqref{part fun}, the critical values of CS-action for $\mathbf{n} \in \mathbb{Z}^{\mathbf{h_1}}$ take the form
\begin{equation}
CS_{\alpha }^{(\mathbf{h_1}, \mathbf{n})} :=
\frac{1}{4} \, \mathbf{n}^T \left( \tilde{A} - \tilde{D}^T \, \tilde{C}^{-1} \, \tilde{D} \right) \mathbf{n}.
\label{eq:S_critical}
\end{equation}

\section{Granny knot with \texorpdfstring{$-1$}{-1}-surgery}\label{plumbed example}
\begin{figure}[t!]
    \centering
    \includegraphics[width=0.8\linewidth]{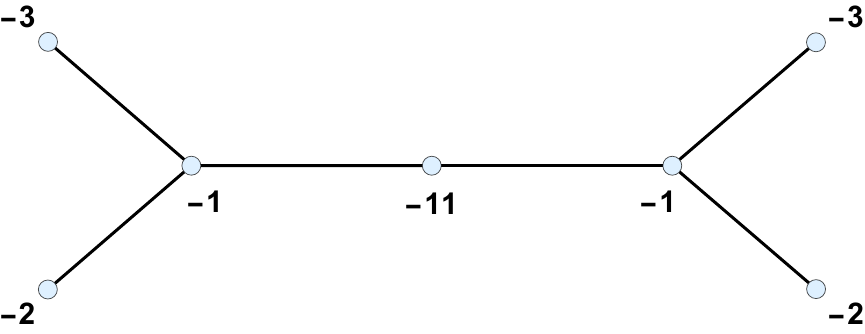}
    \caption{plumbed 3-manifold obtained as $-1$-surgery on the granny knot}
    \label{grannyknot}
\end{figure}

Below we explore the flat connection at infinity for the $-1$-surgery of granny knot using the plumbed graph approach from Appendix \ref{plumbed methd}, detailing the analysis of \cite{gukov2024categorification}. Granny knot is obtained as the connected sum of two right-handed trefoil knots: $3^r_1 \# 3^r_1$.
The plumbed graph presentation of the $-1$-surgery along granny knot is shown in Figure \ref{grannyknot}.
The adjacency matrix reads 
\begin{align}
\mathbf{B}^{(\mathbf{\Gamma})}&=\begin{pmatrix}
-1&0&1&1&1&0&0\\
0&-1&1&0&0&1&1\\
1&1&-11&0&0&0&0\\
1&0&0&-2&0&0&0\\
1&0&0&0&-3&0&0\\
0&1&0&0&0&-3&0\\
0&1&0&0&0&0&-2
\end{pmatrix}
\end{align}
The inverse of  $\mathbf{B}^{(\Gamma)}$ is given by 
\begin{equation}
\mathbf{B^{(\Gamma)}}^{-1} =
\left(\begin{array}{rrrrrrr}
30 & 36 & 6 & 15 & 10 & 12 & 18\\
36&30&6&18&12&10&15\\
6&6&1&3&2&2&3\\
15&18&3&7&5&6&9\\
10&12&2&5&3&4&6\\
12&10&5&6&4&3&5\\
18&15&3&9&6&5&7
\end{array}\right).
\label{binverse}
\end{equation}
Comparing it with the definition given in (\ref{CDA}) we find 
\begin{align}
    \mathbf{C}&=\begin{pmatrix}
        30&36\\
        36&30
    \end{pmatrix},~~~\mathbf{D}=\begin{pmatrix}
        15&10&12&18\\
        18&12&10&15
    \end{pmatrix}.
\end{align}
The explicit form of the functions \eqref{eq:S_function} and \eqref{eq:R_function} are as follows
\begin{equation}
      S(\mathbf{v})=\frac{1}{4}\left( \frac{-5v_1^2 + 12 v_1v_2 -5 v_2^2 }{66}\right)\,,\qquad R(\mathbf{v})=\frac{\sin{\frac{\pi v_1}{2}}\sin{\frac{\pi v_1}{3}}\sin{\frac{\pi v_2}{2}}\sin{\frac{\pi v_2}{3}}}{\sin{\pi v_1}\sin{\pi v_2}}
\,.
\end{equation}
We see that the integrand $R(\mathbf{v})$ has singularities at points where 
$v_1 = n_1$, provided $n_1$ is not divisible by 2 and 3. 
Likewise, singularities occur at $v_2 = n_2$ under the same 
divisibility condition on $n_2$, with $n_1, n_2 \in \mathbb{Z}$. 

To compute the critical values of the Chern-Simons action, we consider subsets of the high-valency vertices. In the case of the granny knot, there are two high-valency vertices, leading to four possible subsets: the empty set $\varnothing$, the  sets $\{1\}$ and $\{2\}$, and the set $\{1, 2\}$. The critical value associated with the empty subset is simply zero. For the subset $\mathbf{h}_1 = \{1\}$, the Chern-Simons action is evaluated to be $CS_\alpha = \frac{23}{24} \mod 1$. Due to symmetry between the two vertices, the same value of CS-action is obtained for $\mathbf{h}_1 = \{2\}$. 

Finally, for the subset $\mathbf{h}_1 $ being $ \{1, 2\}$, the critical value turns out to be $CS_\alpha* = \frac{11}{12} \mod 1$. In this case the value of Chern-Simons action does not correspond to any usual saddle -- in fact this corresponds to the only flat connection at infinity. This argument can be explicitly verified by following the procedure described in Section \ref{sec: counting}. Starting with the irreducible part of the $A$-polynomial of the granny knot
\begin{equation}
\mathcal{A}_{3_1^r \# 3_1^r}^{\rm irr}(x,y)  = (y+x^{-6})^2 (y-x^{-12})\, ,
\end{equation}
and the Alexander polynomial  
\begin{align}
    \Delta_{3_1^r \# 3_1^r}(x^2)&=\left(x^2 - 1 + x^{-2}\right)^2 ,
\end{align}
we find that CS-action for the flat connection at infinity is exactly equal to $\frac{11}{12}~\mod 1$ . This computation suggests that the flat connection at infinity appears as a singularity in the Borel plane. We stress, however, that further careful analysis is required to gain a more complete information about the contribution of such flat connections to the Chern-Simons path integral.

\bibliographystyle{JHEP}
\bibliography{references.bib}

\providecommand{\href}[2]{#2}\begingroup\raggedright\begin{thebibliography}{10}

\bibitem{gukov2024categorification}
S.~Gukov and P.~Putrov, \emph{{On categorification of Stokes coefficients in Chern-Simons theory}},  \href{https://arxiv.org/abs/2403.12128}{{\ttfamily 2403.12128}}.

\bibitem{Ec81}
J.~\'Ecalle, \emph{{\it Fonctions Resurgentes}}, Publ. Math. Orsay (1981).

\bibitem{Dunne:2016nmc}
G.V.~Dunne and M.~\"Unsal, \emph{{New Nonperturbative Methods in Quantum Field Theory: From Large-N Orbifold Equivalence to Bions and Resurgence}}, \href{https://doi.org/10.1146/annurev-nucl-102115-044755}{\emph{Ann. Rev. Nucl. Part. Sci.} {\bfseries 66} (2016) 245} [\href{https://arxiv.org/abs/1601.03414}{{\ttfamily 1601.03414}}].

\bibitem{ABS19}
I.~Aniceto, G.~Basar and R.~Schiappa, \emph{{A Primer on Resurgent Transseries and Their Asymptotics}}, {\emph{{Physics Reports}} {\bfseries 809} (2019) 1} [\href{https://arxiv.org/abs/1802.10441}{{\ttfamily 1802.10441}}].

\bibitem{AM22}
J.r.E.~Andersen and W.E.~Mistegaard, \emph{Resurgence analysis of quantum invariants of {S}eifert fibered homology spheres}, {\emph{J. Lond. Math. Soc. (2)} {\bfseries 105} (2022) 709}.

\bibitem{Baj21}
Z.~Bajnok, J.~Balog, A.~Hegedus and I.~Vona, \emph{{Instanton effects vs resurgence in the O(3) sigma model}}, \href{https://doi.org/10.1016/j.physletb.2022.137073}{\emph{Phys. Lett. B} {\bfseries 829} (2022) 137073} [\href{https://arxiv.org/abs/2112.11741}{{\ttfamily 2112.11741}}].

\bibitem{Marino:2023epd}
M.~Marino, R.~Miravitllas and T.~Reis, \emph{{On the structure of trans-series in quantum field theory}},  \href{https://arxiv.org/abs/2302.08363}{{\ttfamily 2302.08363}}.

\bibitem{Adams:2025qgj}
G.~Adams, O.~Costin, G.V.~Dunne, S.~Gukov and O.~{\"O}ner, \emph{{$c_{eff}$ from Resurgence at the Stokes Line}},  \href{https://arxiv.org/abs/2508.10112}{{\ttfamily 2508.10112}}.

\bibitem{Gukov:2003na}
S.~Gukov, \emph{{Three-dimensional quantum gravity, Chern-Simons theory, and the A polynomial}}, \href{https://doi.org/10.1007/s00220-005-1312-y}{\emph{Commun. Math. Phys.} {\bfseries 255} (2005) 577} [\href{https://arxiv.org/abs/hep-th/0306165}{{\ttfamily hep-th/0306165}}].

\bibitem{Dimofte:2011ju}
T.~Dimofte, D.~Gaiotto and S.~Gukov, \emph{{Gauge Theories Labelled by Three-Manifolds}}, \href{https://doi.org/10.1007/s00220-013-1863-2}{\emph{Commun. Math. Phys.} {\bfseries 325} (2014) 367} [\href{https://arxiv.org/abs/1108.4389}{{\ttfamily 1108.4389}}].

\bibitem{Chung:2014qpa}
H.-J.~Chung, T.~Dimofte, S.~Gukov and P.~Su\l{}kowski, \emph{{3d-3d Correspondence Revisited}}, \href{https://doi.org/10.1007/JHEP04(2016)140}{\emph{JHEP} {\bfseries 04} (2016) 140} [\href{https://arxiv.org/abs/1405.3663}{{\ttfamily 1405.3663}}].

\bibitem{Chun:2019mal}
S.~Chun, S.~Gukov, S.~Park and N.~Sopenko, \emph{{3d-3d correspondence for mapping tori}}, \href{https://doi.org/10.1007/JHEP09(2020)152}{\emph{JHEP} {\bfseries 09} (2020) 152} [\href{https://arxiv.org/abs/1911.08456}{{\ttfamily 1911.08456}}].

\bibitem{Cheng:2022rqr}
M.C.N.~Cheng, S.~Chun, B.~Feigin, F.~Ferrari, S.~Gukov, S.M.~Harrison et~al., \emph{{3-Manifolds and VOA Characters}}, \href{https://doi.org/10.1007/s00220-023-04889-1}{\emph{Commun. Math. Phys.} {\bfseries 405} (2024) 44} [\href{https://arxiv.org/abs/2201.04640}{{\ttfamily 2201.04640}}].

\bibitem{Gukov:2023cog}
S.~Gukov and M.~Jagadale, \emph{{ceff for 3D \ensuremath{\mathscr{N}}=2 theories}}, \href{https://doi.org/10.1142/S0217751X24460126}{\emph{Int. J. Mod. Phys. A} {\bfseries 39} (2024) 2446012} [\href{https://arxiv.org/abs/2308.05360}{{\ttfamily 2308.05360}}].

\bibitem{Gukov:2016njj}
S.~Gukov, M.~Marino and P.~Putrov, \emph{{Resurgence in complex Chern-Simons theory}},  \href{https://arxiv.org/abs/1605.07615}{{\ttfamily 1605.07615}}.

\bibitem{Costin:2023kla}
O.~Costin, G.V.~Dunne, A.~Gruen and S.~Gukov, \emph{{Going to the Other Side via the Resurgent Bridge}},  \href{https://arxiv.org/abs/2310.12317}{{\ttfamily 2310.12317}}.

\bibitem{Harichurn:2025suf}
S.~Harichurn, M.~Jagadale, D.~Noshchenko and D.~Passaro, \emph{{$c_\text{eff}$ from Surgery and Modularity}},  \href{https://arxiv.org/abs/2508.10087}{{\ttfamily 2508.10087}}.

\bibitem{Adams:2025aad}
G.~Adams, O.~Costin, G.V.~Dunne, S.~Gukov and O.~\"Oner, \emph{{Orientation Reversal and the Chern-Simons Natural Boundary}},  \href{https://arxiv.org/abs/2505.14441}{{\ttfamily 2505.14441}}.

\bibitem{Gukov:2017kmk}
S.~Gukov, D.~Pei, P.~Putrov and C.~Vafa, \emph{{BPS spectra and 3-manifold invariants}}, \href{https://doi.org/10.1142/S0218216520400039}{\emph{J. Knot Theor. Ramifications} {\bfseries 29} (2020) 2040003} [\href{https://arxiv.org/abs/1701.06567}{{\ttfamily 1701.06567}}].

\bibitem{Dunne:2012ae}
G.V.~Dunne and M.~Unsal, \emph{{Resurgence and Trans-series in Quantum Field Theory: The CP(N-1) Model}}, \href{https://doi.org/10.1007/JHEP11(2012)170}{\emph{JHEP} {\bfseries 11} (2012) 170} [\href{https://arxiv.org/abs/1210.2423}{{\ttfamily 1210.2423}}].

\bibitem{Shifman:2014fra}
M.~Shifman, \emph{{Resurgence, operator product expansion, and remarks on renormalons in supersymmetric Yang-Mills theory}}, \href{https://doi.org/10.1134/S1063776115030115}{\emph{J. Exp. Theor. Phys.} {\bfseries 120} (2015) 386} [\href{https://arxiv.org/abs/1411.4004}{{\ttfamily 1411.4004}}].

\bibitem{Boito:2021ulm}
D.~Boito and I.~Caprini, \emph{{Renormalons and hyperasymptotics in QCD}}, \href{https://doi.org/10.1140/epjs/s11734-021-00276-w}{\emph{Eur. Phys. J. ST} {\bfseries 230} (2021) 2561}.

\bibitem{Ma21}
M.~Marino, R.~Miravitllas and T.~Reis, \emph{{New renormalons from analytic trans-series}}, \href{https://doi.org/10.1007/JHEP08(2022)279}{\emph{JHEP} {\bfseries 08} (2022) 279} [\href{https://arxiv.org/abs/2111.11951}{{\ttfamily 2111.11951}}].

\bibitem{Reis:2022tni}
T.~Reis, \emph{{On the resurgence of renormalons in integrable theories}}, Ph.D. thesis, U. Geneva (main), 2022.
\newblock \href{https://arxiv.org/abs/2209.15386}{{\ttfamily 2209.15386}}.

\bibitem{bahri1995variational}
A.~Bahri, Y.~Li and O.~Rey, \emph{On a variational problem with lack of compactness: the topological effect of the critical points at infinity}, {\emph{Calculus of Variations and Partial Differential Equations} {\bfseries 3} (1995) 67}.

\bibitem{durfee1998five}
A.H.~Durfee, \emph{Five definitions of critical point at infinity}, {\emph{Singularities: The Brieskorn Anniversary Volume} (1998) 345}.

\bibitem{bahri2006critical}
A.~Bahri, \emph{Critical points at infinity in the variational calculus},  in \emph{Partial Differential Equations: Proceedings of ELAM VIII, held in Rio de Janeiro, July 14--25, 1986}, pp.~1--29, Springer (2006).

\bibitem{behtash2018critical}
A.~Behtash, G.V.~Dunne, T.~Sch{\"a}fer, T.~Sulejmanpasic and M.~{\"U}nsal, \emph{Critical points at infinity, non-gaussian saddles, and bions}, {\emph{Journal of High Energy Physics} {\bfseries 2018} (2018) 1}.

\bibitem{gukov2021two}
S.~Gukov and C.~Manolescu, \emph{A two-variable series for knot complements}, {\emph{Quantum Topology} {\bfseries 12} (2021) 1}.

\bibitem{cooper1994plane}
D.~Cooper, M.~Culler, H.~Gillet, D.D.~Long and P.B.~Shalen, \emph{Plane curves associated to character varieties of 3-manifolds}, {\emph{Inventiones mathematicae} {\bfseries 118} (1994) 47}.

\bibitem{cooper1998representation}
D.~Cooper and D.D.~Long, \emph{Representation theory and the a-polynomial of a knot}, {\emph{Chaos, Solitons \& Fractals} {\bfseries 9} (1998) 749}.

\bibitem{lickorish1962representation}
W.R.~Lickorish, \emph{A representation of orientable combinatorial 3-manifolds}, {\emph{Annals of Mathematics} {\bfseries 76} (1962) 531}.

\bibitem{kirk1990chern}
P.A.~Kirk and E.P.~Klassen, \emph{Chern-simons invariants of 3-manifolds and representation spaces of knot groups}, {\emph{Mathematische Annalen} {\bfseries 287} (1990) 343}.

\bibitem{culler1983varieties}
M.~Culler and P.B.~Shalen, \emph{Varieties of group representations and splittings of 3-manifolds}, {\emph{Annals of Mathematics} {\bfseries 117} (1983) 109}.

\bibitem{boden2006sl}
H.U.~Boden and C.L.~Curtis, \emph{The $\rm sl(2,\mathbb{C})$ casson invariant for seifert fibered homology spheres and surgeries on twist knots}, {\emph{Journal of Knot Theory and Its Ramifications} {\bfseries 15} (2006) 813}.

\bibitem{Gukov:2019mnk}
S.~Gukov and C.~Manolescu, \emph{{A two-variable series for knot complements}}, \href{https://doi.org/10.4171/qt/145}{\emph{Quantum Topol.} {\bfseries 12} (2021) 1} [\href{https://arxiv.org/abs/1904.06057}{{\ttfamily 1904.06057}}].

\bibitem{Harichurn:2023akp}
S.~Harichurn, \emph{{On the $\Delta_a$ invariants in non-perturbative complex Chern-Simons theory}},  \href{https://arxiv.org/abs/2306.11298}{{\ttfamily 2306.11298}}.

\bibitem{clay2013left}
A.~Clay and L.~Watson, \emph{Left-orderable fundamental groups and dehn surgery}, {\emph{International Mathematics Research Notices} {\bfseries 2013} (2013) 2862}.

\bibitem{hoste2004formula}
J.~Hoste and P.D.~Shanahan, \emph{A formula for the a-polynomial of twist knots}, {\emph{Journal of Knot Theory and Its Ramifications} {\bfseries 13} (2004) 193}.

\bibitem{petersen2015polynomials}
K.~Petersen, \emph{A-polynomials of a family of two-bridge knots}, {\emph{New York J. Math} {\bfseries 21} (2015) 847}.

\bibitem{kirk1993chern}
P.~Kirk and E.~Klassen, \emph{Chern-simons invariants of 3-manifolds decomposed along tori and the circle bundle over the representation space of t 2}, {\emph{Communications in mathematical physics} {\bfseries 153} (1993) 521}.

\end{thebibliography}\endgroup
    \end{document}